\documentclass[journal]{IEEEtran}
\usepackage{graphics} 
\usepackage{epsfig} 
\usepackage{mathptmx} 
\usepackage{times} 
\usepackage{amsmath} 
\usepackage{amssymb}  
\usepackage{upgreek}
\usepackage{latexsym,graphicx,epsf,cite,url,bbm,float,amsbsy}
\usepackage{subcaption}
\usepackage{ifpdf}
\usepackage{epstopdf}
\usepackage{color}
\usepackage[table]{xcolor}
\usepackage{tabularx}
\usepackage{stfloats}
\usepackage{tikz}
\usepackage[percent]{overpic}
\usepackage{arydshln}
\usepackage{multirow}
\usepackage{amsthm}
\usepackage{pifont}
\usepackage{enumitem}
\usepackage{cancel}

\newcommand{\Gauss}{\mathcal{N}}
\newcommand{\universe}[1]{\mathbb{#1}}
\newcommand{\player}[1]{$\mathcal{P}_{#1}$}
\newcommand{\diag}{\mathrm{diag}}

\definecolor{myorange}{rgb}{.984,.392,.016}
\newcommand{\colorcell}{\cellcolor{myorange!20!white}}

\newtheorem{theorem}{Theorem}
\newtheorem{lemma}{Lemma}
\newtheorem{proposition}{Proposition}
\newtheorem{corollary}{Corollary}

\theoremstyle{remark}
\newtheorem{remark}{Remark}

\newtheorem*{definition}{Definition}

\newcommand{\calG}{\mathcal{G}}

\newcommand{\hx}{\hat{x}}

\newcommand{\bU}{\bar{U}}

\newcommand{\bLambda}{\bar{\Lambda}}

\newcommand{\bOmega}{\Omega^o}

\newcommand{\bPsi}{\bar{\Psi}}

\newcommand{\tQ}{\tilde{Q}}

\newcommand{\tx}{\tilde{x}}

\newcommand{\ty}{\tilde{y}}

\newcommand{\teta}{\tilde{\eta}}

\newcommand{\nn}{\nonumber}

\newcommand{\trace}{\mathrm{Tr}}
\newcommand{\rank}{\mathrm{rank}}

\DeclareMathOperator*{\argmin}{argmin\;}

\newcommand{\E}{\mathbb{E}}

\newcommand{\cov}{\mathrm{cov}}

\newcommand{\re}{\pmb{e}}
\newcommand{\rn}{\pmb{\vartheta}}

\newcommand{\rs}{\pmb{s}}

\newcommand{\ru}{\pmb{u}}
\newcommand{\rw}{\pmb{w}}
\newcommand{\rx}{\pmb{x}}
\newcommand{\rv}{\pmb{v}}
\newcommand{\ry}{\pmb{y}}
\newcommand{\rz}{\pmb{z}}

\newcommand{\rhx}{\pmb{\hx}}

\newcommand{\rtx}{\pmb{\tx}}

\newcommand{\rty}{\pmb{\ty}}

\newcommand{\N}{\mathbb{N}} 
\newcommand{\R}{\mathbb{R}} 
\newcommand{\AS}{\mathbb{S}} 

\newcommand{\Eta}{\Upsilon}
\newcommand{\Teta}{\tilde{\Upsilon}}

\begin{document}

\title{Persuasion-based Robust Sensor Design Against Attackers with Unknown Control Objectives}

\author{Muhammed~O.~Sayin and~Tamer~Ba\c{s}ar,~\IEEEmembership{Life~Fellow,~IEEE}
\thanks{This research was supported by the U.S. Office of Naval Research (ONR) MURI grant N00014-16-1-2710.}
\thanks{Muhammed O. Sayin is with Laboratory for Information and Decision Systems at MIT, Cambridge, MA 02139. E-mail: sayin@mit.edu}
\thanks{Tamer Ba\c{s}ar is with the Department of Electrical and Computer Engineering, University of Illinois at Urbana-Champaign, Urbana, IL 61801 USA. E-mail: basar1@illinois.edu}}

\maketitle

\begin{abstract}
In this paper, we introduce a robust sensor design framework to provide ``persuasion-based" defense in stochastic control systems against an unknown type attacker with a control objective exclusive to its type. For effective control, such an attacker's actions depend on its belief on the underlying state of the system. We design a robust ``linear-plus-noise" signaling strategy to encode sensor outputs in order to shape the attacker's belief in a strategic way and correspondingly to persuade the attacker to take actions that lead to minimum damage with respect to the system's objective. The specific model we adopt is a Gauss-Markov process driven by a controller with a (partially) ``unknown" malicious/benign control objective. We seek to defend against the worst possible distribution over control objectives in a robust way under the solution concept of Stackelberg equilibrium, where the sensor is the leader. We show that a necessary and sufficient condition on the covariance matrix of the posterior belief is a certain linear matrix inequality and we provide a closed-form solution for the associated signaling strategy. This enables us to formulate an equivalent tractable problem, indeed a semi-definite program, to compute the robust sensor design strategies ``globally" even though the original optimization problem is non-convex and highly nonlinear. We also extend this result to scenarios where the sensor makes noisy or partial measurements. Finally, we analyze the ensuing performance numerically for various scenarios.
\end{abstract}


\section{Introduction}

\IEEEPARstart{C}{yber} connectedness of control systems has brought in new security challenges where attackers can manipulate control systems at unprecedented levels with various malicious tasks on their agenda\cite{ref:Giraldo17, ref:Humayed17, ref:Nelson16}. We can view such intelligent attackers as decision makers that take actions driven by and compatible with their own objective and information available to them. This implies that the kind of information available to an attacker has an indirect impact on what actions the attacker would take. Correspondingly, it is intuitively expected that if we can manipulate the information available to attackers, then we can {\em persuade} them to attack the system in a way in line with the system's objective to the extent possible, so that the attack would cause minimum damage to the system. 

There are certain distinct challenges for persuasion-based defense measures. For example, attackers make their decisions based on the belief they have formed using the information available to them. A challenge is how to shape that belief in a desired way by controlling the information available to the attacker. The first requirement (and challenge) here is to be able to identify how an attacker would form its belief. However, when an attacker forms its belief strategically, there might be multiple Nash equilibria in a general-sum setting\footnote{In a zero-sum setting, there exists a unique ``babbling equilibrium" where the attacker forms a belief independent of the information provided by a defender (since the defender would not share anything useful) whereas the defender just makes irrelevant information available to the attacker (since a more informed attacker would not cause less damage to the system).}, as shown in the strategic information transmission framework \cite{ref:Crawford82}. In that case how an attacker forms its belief is not well defined since the attacker might be forming its belief differently at different equilibria and whether (and which) equilibrium will be realized would not be known, and constitutes an uncertainty. Furthermore, even when there exists a well-defined characterization of how an attacker would form its belief, another challenge is what decision the attacker would make based on that belief, since that decision would depend on the attacker's malicious objective. This implies that persuasion-based defense is attack-specific. Therefore another challenge is how to persuade an attacker whose objective is not known to act in a certain way. 

Before addressing these challenges, let us briefly review the literature from a broader perspective where a decision maker seeks to induce another one to take certain actions. A closely related framework is Bayesian persuasion, introduced by Kamenica and Gentzkow in their seminal paper \cite{ref:Kamenica11}. They addressed how a sender could persuade a receiver to take certain actions under the solution concept of Stackelberg equilibrium \cite{ref:Basar99} where the sender (the receiver) is the leader (the follower). In other words, the receiver is aware of the sender's signaling strategy while taking its actions. Such a leader-follower scheme creates an environment where the receiver's actions are not strategic, and therefore there is a well-defined characterization for how the belief is formed. 

In Bayesian persuasion framework, a necessary and sufficient condition is that mean of posterior belief must be equal to the prior belief. This enables us to formulate an equivalent problem over distributions over posterior beliefs under a linear equality constraint. The authors provided a geometrical interpretation to compute the solution, which necessitates computation of a convex envelope of a function. Although this interpretation provides essential insight for designing signaling strategies and has led to various applications (see the recent survey on Bayesian persuasion \cite{ref:Kamenica19} and the references therein), it is viable only for scenarios where the state space is very small. 

We emphasize that \cite{ref:Kamenica11} studies the Bayesian persuasion problem in a very general framework, where the underlying distribution is arbitrary as long as its support set is compact, cost functions are arbitrary, and the signaling strategy is any stochastic kernel (between the state space and the signal space). Therefore if we consider specific distributions and cost functions, we should be able to obtain more tractable solutions. For example in \cite{ref:Tamura14}, the author addressed Bayesian persuasion problem for multi-variate Gaussian information and quadratic cost functions, and provided a closed-form solution for the optimal signaling strategy, which turns out to be a linear function. We note that the studies \cite{ref:Kamenica11,ref:Tamura14} focused on static systems. To be able to adopt this framework to control systems, an important step would be to extend the framework to dynamic systems. 
In \cite{ref:Sayin17b}, we have extended the results in \cite{ref:Tamura14} to dynamic environments where the underlying information is a discrete-time Gauss Markov process. We showed that there exists a linear signaling strategy optimal within the general class of measurable policies and provided a semi-definite program (SDP) to compute optimal signaling strategies numerically.

Before delving into persuasion-based defense measures in control systems, let us also review the literature on security of control systems. To this end, we selectively focus on studies where an attacker can monitor and intervene the links from sensor to controller and from controller to actuator so that there would be an information flow to the attacker who monitors these links. In \cite{ref:Liu09}, the authors showed that an attacker can lead to unbounded estimation error by injecting false data into the link from sensor to controller when there exists a certain threshold-based detector monitoring the link. In \cite{ref:Mo12,ref:Mo16}, the authors characterized the reachable set to which an evasive attacker can drive the system by injecting data into both links jointly. In \cite{ref:Chen16a,ref:Chen16b}, the authors analyzed attacks where an attacker seeks to drive the state of the system according to his/her adversarial goal evasively by injecting data into both links jointly. In \cite{ref:Zhang17}, the authors have analyzed optimal attack strategies to maximize quadratic cost of a system with linear Gaussian dynamics by injecting false data into the link from controller to actuator without being detected. 

\begin{figure}[t!]
\begin{center}
\begin{overpic}[width=.5\textwidth]{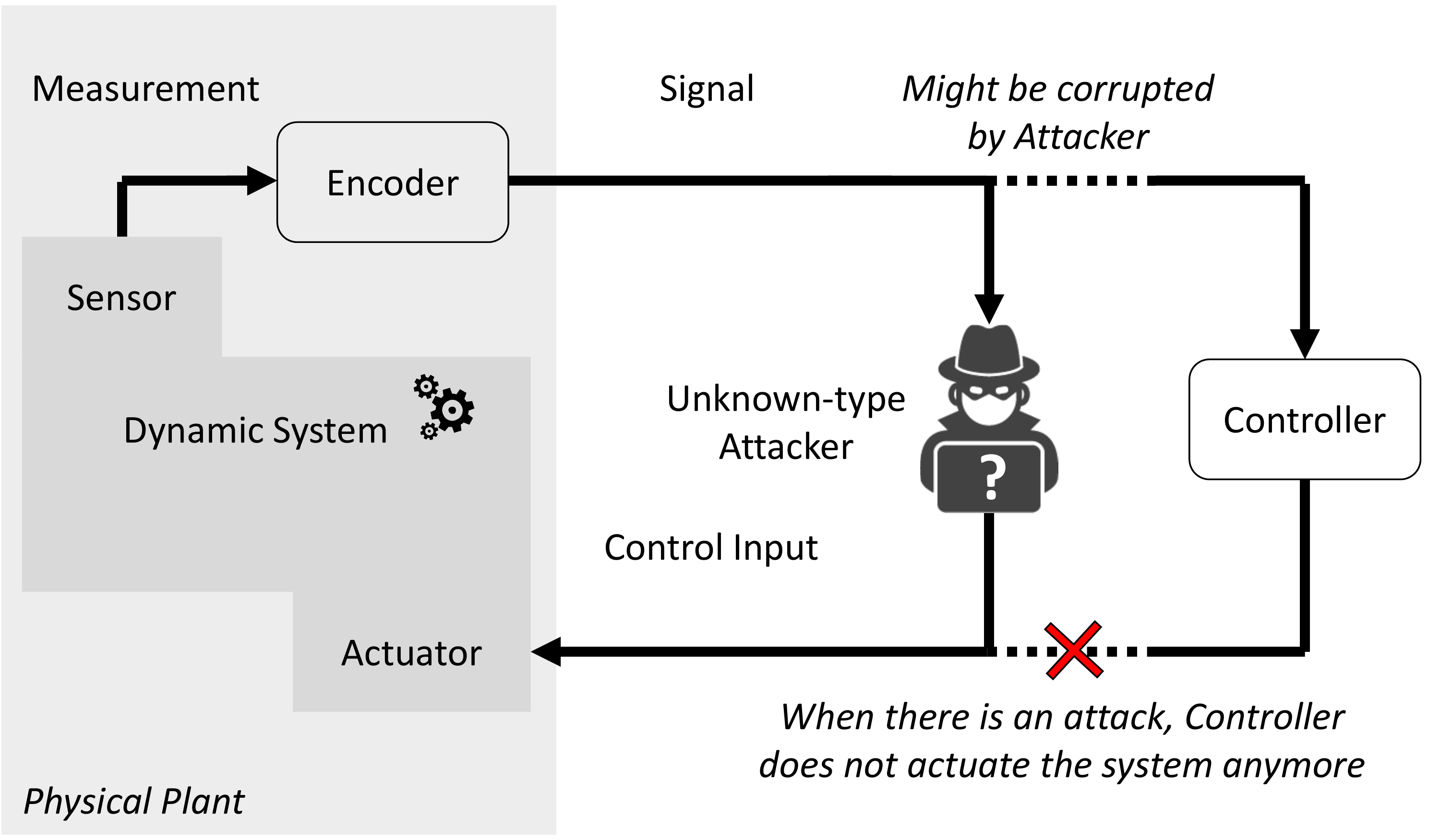}
\put(3,22){\small $\rx_{k+1} = A\rx_k + B\ru_k + \rw_k$}
\put(41.5,16){\small $\ru_k= \gamma_k(\rs_{1:k})$}
\put(8,48){\small $\ry_k$}
\put(24,38) {\small $\ry_{1:k-1}$}
\put(41,48){\small $\rs_k = \eta_k(\ry_{1:k})$}
\end{overpic}
\end{center}
\caption{A discrete-time LQG control system where an (unknown-type) attacker can monitor and intervene the links from sensor to controller, and from controller to actuator jointly in order to drive the system according to its malicious control objective. The attacker might also corrupt the data received by the controller of the system to evade a detector monitoring that data. Details of how an attacker can evade detection by corrupting that data has earlier been studied in the literature, e.g., \cite{ref:Liu09, ref:Mo12, ref:Mo16, ref:Chen16a, ref:Chen16b}, and is out of the scope of this paper since here we focus on adopting persuasion-based defense against attackers with unknown control objectives. Furthermore, the results of this paper hold for any strategy that the attacker can use to evade such detectors as long as those strategies do not impact how the attacker constructs its control inputs.}\label{fig:model}
\end{figure}
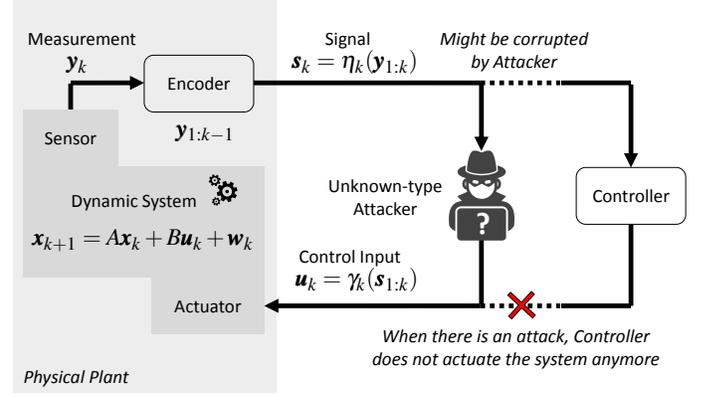

As seen in the literature reviewed above, an attacker can monitor and intervene the links in a control system, e.g., as illustrated in Fig. \ref{fig:model} for a linear quadratic Gaussian (LQG) control system. This implies that an attacker can bypass the controller of a system by monitoring and intervening both links jointly. Then the attacker would receive the sensor outputs and could generate its own control inputs to conduct its malicious control objective similar to the controller of the system. From this viewpoint, within the Bayesian persuasion framework, we can encode the sensor outputs to persuade the attacker to generate control inputs that would minimize the system's very own control objective as much as possible. We partially addressed this challenge in \cite{ref:Sayin17b, ref:Sayin17a, ref:Sayin18a}. In \cite{ref:Sayin17b}, we formulated an optimal linear signaling rule in an LQG setting when attacker's control objective is known to the sensor. This limits applicability of the solution since the sensor must know when there is an attack and what the control objective of the attacker is beforehand. In the follow-up papers \cite{ref:Sayin17a, ref:Sayin18a}, we showed that this limitation could be relaxed in a straightforward way if the sensor knows the distribution over the control objectives of attackers that may attack the system and at what probability there might be an attack. Correspondingly, a risk-neutral sensor could minimize the expected damage with respect to that distribution. However, this still limits its applicability since the sensor must know the underlying distribution over the attack space. 

Although it is not a persuasion-based defense, it is worth noting that in \cite{ref:Miao17}, the authors proposed linear encoding schemes\footnote{We use the terminologies encoding scheme and signaling rule/strategy interchangeably.} for sensor outputs in an LQG control problem in order to enhance detectability of false data injection attacks under an essential assumption that the encoding matrix is oblivious to attackers. In that respect, the encoding scheme could be viewed as corrupting the information available to attackers (without impacting the information flowing to the controller of the system) in order to limit the attacker's capability to evade detection rather than persuading the attacker to attack in a certain way. And this security measure becomes undermined completely once the encoding scheme is compromised.

Coming to the specifics of this paper, we also consider a discrete-time LQG control problem, similar to the studies reviewed above \cite{ref:Chen16a,ref:Chen16b, ref:Zhang17, ref:Sayin17a, ref:Sayin18a, ref:Miao17}, but with important differences. We particularly seek to design {\em linear-plus-noise} signaling rules for persuasion-based robust sensor design against an unknown type attacker with a control objective exclusive to its type. It is robust in the sense that sensor outputs are designed against the worst possible distribution over all possible control objectives of the attacker. We consider the scenario where the set of types is finite and known to the system. Under the solution concept of Stackelberg equilibrium, we consider the setting where the sensor is the leader and the attacker is the follower. This yields that the underlying encoding scheme is not necessarily oblivious to the attacker. Furthermore, we address the scenarios where the sensor can have partial or noisy measurements different from the models in \cite{ref:Sayin17b,ref:Sayin17a, ref:Sayin18a}. Note that the worst case distribution over the type set turns out to be not necessarily a degenerate distribution, i.e., defending only against the (strongest) type of attack that leads to largest cost for the system is not necessarily optimal with respect to the system's objective. 

Due to the underlying leader-follower setting, the response of the follower (the attacker) is non-strategic and the problem faced by the leader (the sensor) turns out to be an optimization problem rather than a fixed-point problem. By using the classical method of completion of squares \cite{ref:Kumar86}, we can show that the objective function in that optimization problem is linear in the covariance matrix of the posterior estimate of the underlying (control-free) state. However that covariance matrix depends on the encoding scheme in a nonlinear way, which leads to a {\em non-convex} and {\em highly nonlinear} optimization problem, to be solved {\em globally} in order to compute robust sensor design strategies.  In \cite{ref:Sayin17b}, our previous inspection revealed a necessary condition on this covariance matrix, which is just a linear matrix inequality. Here we show that this necessary condition is also a {\em sufficient} condition when the sensor selects linear plus noise signaling strategies. Furthermore for any matrix satisfying the necessary condition, we provide a closed-form solution for the associated linear plus noise signaling rule. Therefore instead of trying to solve a non-convex and highly nonlinear optimization problem, we are able to bring the problem to one of solving a linear optimization problem under linear matrix inequality constraints, which can be done numerically using existing SDP solvers effectively, e.g., \cite{ref:CVX,ref:Grant08}. This solution concept can be seen to have similar flavor with our previous paper \cite{ref:Sayin17b}, reviewed above. However, here we develop and present a completely new and more comprehensive set of technical tools since the results of \cite{ref:Sayin17b} cannot be adopted for the general setting of this paper. The reader can refer to Appendix \ref{app:comparison} for a detailed discussion on this matter.

We now highlight the main contributions of this paper as follows:
\begin{itemize}[leftmargin=.2in]
\item We show that a {\em necessary and sufficient} condition on the covariance matrix of the posterior estimate of control-free state is a certain linear matrix inequality. This result is important by itself since it yields a tractable solution concept to design sensor outputs in general settings not limited to the special setting studied in this paper. 
\item Based on this necessary and sufficient condition, we provide an SDP {\em equivalent} to the original optimization problem faced by the sensor, which is non-convex and highly nonlinear, while the SDP could be solved globally using existing powerful computational tools effectively \cite{ref:CVX,ref:Grant08}. Furthermore, this result can be extended to scenarios where there are partial or noisy measurements.
\item We note that robust signaling rule can dictate sensors to introduce irrelevant information into sensor outputs quite contrary to  other settings, e.g., when there is no uncertainty (or there is imperfect information) on attacker's type. This observation enables us to draw the following conclusions:
\begin{itemize}
\item The equivalence result does not necessarily hold if the sensor can only use linear signaling rules.
\item Based on Blackwell's Irrelevant Information Theorem \cite[Theorem D.1.1]{ref:Yuksel13}, linear signaling strategies are not the best one within the general class of measurable strategies in this robust setting.
\end{itemize}
\end{itemize}

The paper is organized as follows: In Section \ref{sec:prob}, we formulate the robust sensor design game. In Section \ref{sec:robust}, we analyze the equilibrium of the robust sensor design game under perfect measurements. In Section \ref{sec:noisy}, we extend the results to the cases where there are partial or noisy measurements.  In Section \ref{sec:examples}, we examine numerically the performance of the proposed scheme for various scenarios. We conclude the paper in Section \ref{sec:conclusion} with several remarks and possible research directions. An appendix provides further discussion on related literature, proofs of all technical results in the order they appear throughout the paper, and some closed-form expressions for the reader's reference.

{\em Notation:} We denote a collection of parameters via a subscript, e.g., $\{x_k\}$, by dropping the subscript, e.g., $x$, for notational brevity. For a vector $x$ and a matrix $A$, $x'$ and $A'$ denote their transposes, and $\|x\|$ denotes the Euclidean ($L^2$) norm of the vector $x$. For a matrix $A$, $\trace\{A\}$ and $\rank\{A\}$ denote its trace and rank, respectively. We denote the identity and zero matrices with the associated dimensions by $I$ and $O$, respectively. We denote the set of $m$-by-$m$ symmetric, positive semi-definite, and positive definite matrices by $\universe{S}^m$, $\universe{S}_+^m$, and $\universe{S}_{++}^m$, respectively. For a matrix $A$, $A^{\dagger}$ denotes its Moore-Penrose inverse. For positive semi-definite matrices $A$ and $B$, $A\succeq B$ means that $A-B$ is also a positive semi-definite matrix. We denote the Kronecker product of matrices $A$ and $B$ by $A\otimes B$.

We denote an ordered set $\{x_k,\ldots,x_1\}$ and its augmented vector version, $\begin{bmatrix} x_k' & \cdots & x_1' \end{bmatrix}'$, by $x_{1:k}$, by some abuse of notation. $\Gauss(0,.)$ denotes the multivariate Gaussian distribution with zero mean and designated covariance matrix. We denote random variables by bold lower case letters, e.g., $\rx$. We denote expectation and (co)variance of a random variable $\rx$ by $\E\{\rx\}$ and $\cov\{\rx\}$, respectively. For random variables $\rx$ and $\ry$, $\E\{\rx|\ry\}$ denotes the expectation of $\rx$ with respect to the random variable $\ry$. We denote the set of all possible probability distributions over a set $\Omega$ by $\Delta(\Omega)$.

\section{Problem Formulation}\label{sec:prob}

Consider a control system, as illustrated in Fig. \ref{fig:model}, whose underlying state dynamics and sensor measurements are described, respectively, by:
\begin{align}
&\rx_{k+1} = A\rx_k + B\ru_k + \rw_k, \label{eq:state}\\
&\ry_k = C\rx_k + \rv_k, \label{eq:measurement}
\end{align}
for $k=1,\ldots,\kappa$, where\footnote{Even though we consider time-invariant matrices $A, B$, and $C$, for notational simplicity, the provided results could be extended to time-variant cases rather routinely. Furthermore, we consider all the random parameters to have zero mean; however, the derivations can be extended to non-zero mean case in a straight-forward way.} $A\in\R^{m\times m}, B\in\R^{m\times r}$, and $C\in\R^{n\times m}$, and the initial state $\rx_1\sim\Gauss(0,\Sigma_1)$. The additive state and measurement noise sequences $\{\rw_k\}$ and $\{\rv_k\}$, respectively, are white Gaussian vector processes, i.e., $\rw_k \sim \Gauss(0,\Sigma_w)$ and $\rv_k\sim\Gauss(0,\Sigma_v)$; and are independent of the initial state $\rx_1$ and of each other. 

As seen in Fig. \ref{fig:model}, measurements $\ry_{1:k}\in\R^{nk}$ are encoded into a signal $\rs_k\in\R^m$ through a signaling strategy $\eta_k(\cdot)$, which is a stochastic kernel, and $\rs_k = \eta_k(\ry_{1:k})$ almost everywhere over $\R^m$. We specifically consider ``linear plus noise" signaling rules such that the signal $\rs_k$ is given by
\begin{align}\label{eq:signal}
\rs_k &= \eta_k(\ry_{1:k}) = L_{k}'\ry_{1:k} + \rn_k,
\end{align}
almost everywhere over $\R^m$, where $L_{k}\in\R^{nk\times m}$ can be any $nk$-by-$m$ deterministic matrix, and $\rn_k\sim\Gauss (0,\Theta_k)$ is a zero mean multivariate Gaussian noise independent of every other parameter, and its covariance matrix $\Theta_k\in\universe{S}^m_+$ can be any $m$-by-$m$ positive semi-definite matrix. We denote the set of such signaling rules by $\Eta_k$, i.e., $\eta_k\in\Eta_k$. Furthermore, the closed-loop control input $\ru_k\in\R^r$ is given by 
\begin{equation}\label{eq:control}
\ru_k = \gamma_k(\rs_{1:k}),
\end{equation}
almost everywhere over $\R^r$, where $\gamma_k(\cdot)$ can be any Borel measurable function from $\R^{mk}$ to $\R^r$. We denote the set of such control policies by $\Gamma_k$, i.e., $\gamma_k\in\Gamma_k$. For notational brevity, let us denote signaling (control) strategies and the associated sets across the horizon by $\eta$ and $\Eta$ ($\gamma$ and $\Gamma$), respectively.

\subsection{Defense Model}

We consider an LQG control problem, where the controller of the system selects a measurable control strategy $\gamma\in\Gamma$ in order to minimize 
\begin{equation}\label{eq:Cobj}
\sum_{k=1}^{\kappa} \E\|\rx_{k+1}\|_Q^2 + \E\|\ru_k\|_R^2,
\end{equation}
where\footnote{For notational simplicity, we consider time-invariant $Q$ and $R$. However, the results provided could be extended to the general time-variant case rather routinely.} $Q\in\universe{S}^m_+$ and $R\in\universe{S}_{++}^r$. As illustrated in Fig. \ref{fig:model}, we consider the encoder of the system as a decision maker, denoted by \player{S}. \player{S} selects the signaling strategy $\eta\in\Eta$ in order to minimize the same objective with the controller, \eqref{eq:Cobj}. 
Note that if there were no attacks, identity function, where the measurements are shared with the controller fully, would be an optimal encoding scheme due to the data processing inequality \cite[Theorem 2.8.1]{ref:Cover06}. However, we consider here the scenarios where there can be an attack with an unknown control objective. Correspondingly there might be encoding schemes that do not share the measurements fully and can lead to better performance with respect to \eqref{eq:Cobj}.

\subsection{Attack Model}

We consider an attacker who is aware of the underlying state dynamics, i.e., gain matrices $A$, $B$, and $C$; covariance matrices $\Sigma_1$, $\Sigma_w$, and $\Sigma_v$; and the encoding scheme, i.e., $\eta$. The attacker is of an unknown type, which determines its control objective. Let us denote the set of all possible types by $\Omega$. We suppose that $\Omega$ is {\em finite} and {\em known} by the system. We seek to provide a compact and unified analysis. Therefore we consider that the control objective of type $\omega\in\Omega$ is given by
\begin{equation}\label{eq:AAobj}
\sum_{k=1}^{\kappa} \E\|\rx_{k+1}^{\omega}\|_{Q^{\omega}}^2 + \E\|\ru_k^{\omega}\|_{R^{\omega}}^2,
\end{equation}
where $Q^{\omega}\in\universe{S}_+^m$ and $R^{\omega}\in\universe{S}_{++}^r$ are exclusive to the type $\omega$. Note that when there is an attack, the attacker selects a control strategy $\gamma^{\omega}\in\Gamma$ in order to construct its control input $\ru_k^{\omega}$ and correspondingly its control objective includes the term $\E\|\ru_k^{\omega}\|_{R^{\omega}}^2$. We also denote the state driven by type-$\omega$ attacker by $\rx_{k}^{\omega}$, to make it explicit. 

\begin{remark}[Versatility of Control Objectives]
We model the control objectives of the system and the attacker by \eqref{eq:Cobj} and \eqref{eq:AAobj}, respectively, within a unified framework. However, arbitrariness of weight matrices in the control objectives \eqref{eq:Cobj} and \eqref{eq:AAobj} brings in flexibility to model various attack scenarios (that are not in the exact form of \eqref{eq:AAobj}) through the transformation of the underlying state space as exemplified in Section \ref{sec:examples}.\hfill $\square$
\end{remark} 

\subsection{Game Model}

We analyze the interaction between the attacker and \player{S} under the solution concept of Stackelberg equilibrium where \player{S} is the leader. Note that from the viewpoint of \player{S}, either the controller of the system is receiving the sensor output and driving the state or there is an unknown type attack, and it is getting the sensor output and it is generating the control input. Since whether there is an attack or not is also an uncertainty, let us view the attacker and the controller of the system as a single player in a unified way, denoted by \player{C}, with an unknown type from the extended type set $\bOmega := \Omega \cup \{\omega_o\}$, where type-$\omega_o$ corresponds to the controller of the system, i.e., $Q^{\omega_o} = Q$ and $R^{\omega_o} = R$. Therefore, we consider a Stackelberg game between the leader \player{S} and the follower \player{C} (of an unknown type).

We note that depending on its type, \player{C} selects different control policies, which lead to different control inputs, and states. Therefore for type-$\omega$ \player{C}, we use $\gamma_{k}^{\omega}$, $\rx_k^{\omega}$ and $\ru_k^{\omega}$. The objective of type-$\omega$ \player{C} is given by
\begin{equation}
U_{C}^{\omega}(\eta,\gamma^{\omega}) := \sum_{k=1}^{\kappa}\E\|\rx_{k+1}^{\omega}\|_{Q^{\omega}}^2 + \E\|\ru_k^{\omega}\|_{R^{\omega}}^2.\label{eq:typeCobj}
\end{equation} 
On the other hand, the objective of \player{S} is given by
\begin{align}
U_{S}(\eta,\{\gamma^{\omega}\}_{\omega\in\bOmega}) &:= \max_{p \in \Delta(\bOmega)} \Bigg\{\sum_{\omega\in\Omega} p_{\omega}\sum_{k'=1}^{\kappa} \E\|\rx_{k'+1}^{\omega}\|_{Q^{\omega_o}}^2\nn\\
+& p_{\omega_o}\sum_{k=1}^{\kappa}\left(\E\|\rx_{k+1}^{\omega_o}\|_{Q^{\omega_o}}^2 + \E\|\ru_k^{\omega_o}\|_{R^{\omega_o}}^2\right)\Bigg\},\label{eq:Sobj}
\end{align}
where $\Delta(\bOmega)$ denotes all possible distributions over the extended type set $\bOmega$. Note that the maximization in \eqref{eq:Sobj} computes the cost of \player{S} for the worst case distribution over $\bOmega$.

Before describing the game formally, let us take a closer look at \player{S}'s objective \eqref{eq:Sobj}. We can view \eqref{eq:Sobj} consisting of two parts, one of which is
\begin{equation}\label{eq:first}
p_{\omega_o}\sum_{k=1}^{\kappa}\left(\E\|\rx_{k+1}^{\omega_o}\|_{Q^{\omega_o}}^2 + \E\|\ru_k^{\omega_o}\|_{R^{\omega_o}}^2\right),
\end{equation}
where $p_{\omega_o}$ corresponds to the probability that the controller of the system drives the state under the worst case distribution and the summation is identical to \eqref{eq:Cobj} since $Q^{\omega_o} = Q$ and $R^{\omega_o} = R$. The other part is 
\begin{equation}\label{eq:second}
 \sum_{\omega\in\Omega} p_{\omega}\sum_{k'=1}^{\kappa} \E\|\rx_{k'+1}^{\omega}\|_{Q^{\omega_o}}^2,
\end{equation}
which implies that \player{S} seeks to minimize $\sum_{k=1}^{\kappa} \E\|\rx_{k+1}^{\omega}\|_{Q^{\omega_o}}^2$ when type-$\omega$ attacker drives the state. Note that it includes only the first term in \eqref{eq:Cobj} since we consider that \player{S} would not necessarily want the attacker to have small size control inputs.  

\begin{definition}[Robust Sensor Design Game]
The robust sensor design game
\begin{equation}
\calG := \left(\Eta,\Gamma,\rx_1,\rw_{1:\kappa},\rv_{1:\kappa}\right)
\end{equation}
is a Stackelberg game between the leader \player{S} and the follower \player{C}. Objectives of \player{C} and \player{S} are given by \eqref{eq:typeCobj} and \eqref{eq:Sobj}, respectively. Then a tuple of strategies $(\eta,\{\gamma^{\omega}\}_{\omega\in\bOmega})$ attains the Stackelberg equilibrium provided that 
\begin{subequations}\label{eq:equilibrium}
\begin{align}
&\eta \in \argmin_{\eta'\in\Eta} U_{S}\big(\eta',\{\gamma^{\omega}(\eta')\}_{\omega\in\bOmega}\big)\label{eq:etaStar}\\
&\gamma^{\omega}(\eta) \in \argmin_{\gamma\in\Gamma} U_{C}^{\omega}\big(\eta,\gamma(\eta)\big).\label{eq:gammaStar}
\end{align}
\end{subequations}
where, by an abuse of notation, we denote type-$\omega$ \player{C}'s strategy $\gamma^{\omega}$ by $\gamma^{\omega}(\eta)$ to show its dependence on \player{S}'s signaling rules due to the leader-follower scheme, explicitly. 
\end{definition}

Note that there might be multiple best responses by \player{C} for a signaling strategy. Correspondingly \eqref{eq:etaStar} would not be well defined if these best responses lead to different costs for \player{S}. However, as we will show later in detail, reaction set of type-$\omega$ \player{C} turns out to be an equivalence class such that all $\gamma^{\omega}$ in the reaction set lead to the same control input almost surely, and correspondingly lead to the same cost for \player{S}.


\section{Robust Sensor Design Framework}\label{sec:robust}

In this section, we consider the special case where \player{S} has access to perfect measurements, i.e., $\ry_k = \rx_k$ for $k=1,\ldots,\kappa$; the general noisy/partial measurements case will be addressed later in Section \ref{sec:noisy}. To compute the equilibrium of the game $\calG$, we first focus on best response strategy of \player{C} for a given signaling strategy. This enables us to formulate the optimization problem faced by \player{S} to compute robust signaling strategies. Even though this is a finite-dimensional optimization problem, further inspection reveals that it is highly nonlinear and non-convex. Therefore a generic approach would not be able to address it globally. To mitigate this issue, we formulate a tractable problem equivalent to the original optimization problem. We can solve this tractable problem globally using existing computational tools effectively. Given that solution, we also show how we can compute the associated signaling strategies. We now provide the details of these steps.

An important challenge in the design of encoding schemes in control systems compared to communication systems is that the underlying state depends on control inputs. To mitigate this issue, we introduce control-free state $\{\rx_k^o\}$ evolving according to
\begin{equation}\label{eq:free}
\rx_{k+1}^o = A \rx_k^o + \rw_k,
\end{equation}
and $\rx_1^o = \rx_{1}$. As shown in \cite{ref:Bansal89}, the routine technique of completion of squares yields that  
\begin{equation}
\sum_{k=1}^{\kappa} \E\|\rx_{k+1}\|_{Q^{\omega}}^2 + \E\|\ru_k\|_{R^{\omega}}^2 = \sum_{k=1}^{\kappa} \E\|\ru_k^o + K_k^{\omega} \rx_k^o\|_{\Delta_k^{\omega}}^2 + \delta_0^{\omega},\label{eq:control-free}
\end{equation}
where the matrices $K_k^{\omega}$, $\Delta_k^{\omega}$, and scalar $\delta_o^{\omega}$ are given by
\begin{align*}
&K_{k}^{\omega}=(\Delta_{k}^{\omega})^{-1}B'\tQ_{k+1}^{\omega}A\\
&\Delta_{k}^{\omega} = B'\tQ_{k+1}^{\omega}B + R^{\omega}\\
&\delta_{0}^{\omega} = \trace\{Q^{\omega}\Sigma_1\} + \sum_{k=1}^{\kappa}\trace\{\tQ_{k+1}^{\omega}\Sigma_w\}
\end{align*}
and $\{\tQ_{k}^{\omega}\}$ satisfies the following discrete-time dynamic Riccati equation
\begin{equation*}
\tQ_{k}^{\omega} = Q^{\omega} + A'(\tQ_{k+1}^{\omega} - \tQ_{k+1}^{\omega}B(\Delta_{k}^{\omega})^{-1}B'\tQ_{k+1}^{\omega})A
\end{equation*}
and $\tQ_{\kappa+1}^{\omega} = Q^{\omega}$. Furthermore $\ru_k^o$ depends on the control inputs through the following transformation:
\begin{align*}
\ru_k^o = \ru_k + K_k^{\omega}B\ru_{k-1} + \ldots + K_k^{\omega}A^{k-2}B\ru_1.
\end{align*}
Note that $\Delta_k^{\omega}$ is positive definite for all $k$.

Contrary to team problems (where all decision makers have the same objective), as studied in \cite{ref:Bansal89}, in non cooperative settings, \eqref{eq:control-free} does not imply that a control input leading to $\ru_k^o = - K_k^{\omega}\E\{\rx_k^o|\rs_{1:k}\}$ is optimal since control inputs can have an impact on the signals that will be generated in future stages. However, as we will show below, \player{C} cannot impact the signals generated in future stages when \player{S} uses linear plus noise signaling strategies only. To show this, we let the gain matrix $L_k\in\R^{nk\times m}$ in signaling strategy $\eta_k$, as described in \eqref{eq:signal}, be partitioned as $L_{k} = \begin{bmatrix} L_{k,k}' & \cdots & L_{k,1}' \end{bmatrix}'$, where $L_{k,j} \in \R^{n\times m}$. Then, signal $\rs_k$ can be written as
\begin{align*}
L_{k,k}'\rx_k + &\ldots + L_{k,1}'\rx_{1} + \rn_k = L_{k,k}'\rx_k^o + \ldots + L_{k,1}'\rx_{1}^o + \rn_k \nn \\
+& \Big\{L_{k,k}'B\ru_{k-1}+\ldots+(L_{k,k}'A^{k-2} + \ldots + L_{k,2}')B\ru_{1}\Big\},
\end{align*}
where the term in-between $\{\cdot\}$ is $\sigma$-$\rs_{1:k-1}$ measurable. Similar to\footnote{We note that \cite[Lemma 12]{ref:Sayin17b} shows a result similar to \eqref{eq:optU} when the sender selects only linear and memoryless signaling strategies.} \cite[Lemma 12]{ref:Sayin17b}, this yields that
\begin{align}\label{eq:so}
\E\{\rx_k^o | \rs_{1:k}\} = \E\{\rx_k^o|\rs_{1:k}^o\},
\end{align}
where we define $\rs_k^o := L_{k}'\rx^o_{1:k}+\rn_k$. Correspondingly, since $\Delta_k^{\omega}$ is positive definite for all $k$, right-hand sides of \eqref{eq:control-free} and \eqref{eq:so} yield that optimal reaction of type-$\omega$ \player{C} is indeed the one that leads to
\begin{equation}\label{eq:optU}
\ru_k^o = - K_k^{\omega}\E\{\rx_k^o|\rs_{1:k}\},
\end{equation}
almost everywhere over $\R^r$, since $\E\{\rx_k^o|\rs_{1:k}\}$ does not depend on $\ru_{1:k}$. This shows that all strategies in the best reaction set of type-$\omega$ \player{C} lead to \eqref{eq:optU} and therefore lead to the same cost for \player{S}.

Based on \eqref{eq:optU}, the following lemma shows that we can write the optimization objective in \eqref{eq:Sobj} as a linear function of the covariance matrix of the posterior estimate of control-free state, i.e.,\footnote{Henceforth we say ``covariance matrix" instead of ``covariance matrix of posterior estimate of control-free state", and we say ``posterior" instead of ``posterior estimate of control-free state".}
\begin{equation*}
H_k := \cov\{\E\{\rx_k^o|\rs_{1:k}\}\},
\end{equation*}
for $k=1,\ldots,\kappa$.

\begin{lemma}\label{lem:linear}
The problem faced by \player{S}, i.e., \eqref{eq:etaStar}, can be written as
\begin{equation}\label{eq:short}
\min_{\eta\in\Eta}\max_{p\in\Delta(\bOmega)} \sum_{\omega\in\bOmega}p_{\omega}\left(\xi^{\omega} + \sum_{k=1}^{\kappa}\trace\left\{H_k\Xi_k^{\omega}\right\}\right),
\end{equation}
where $\Xi_{k}^{\omega}$ is a certain symmetric matrix, described in \eqref{eq:V}, that does not depend on the optimization arguments, and $\xi^{\omega}\in\R$, described in \eqref{eq:xi1} and \eqref{eq:xi2}, is a certain fixed scalar. 
\end{lemma}

We emphasize that complexity of the objective functions \eqref{eq:Cobj} and \eqref{eq:Sobj} is buried in fixed parameters $\{\{\Xi_k^{\omega}\},\xi^{\omega}\}$. Based on this observation, we make the following remarks:

\begin{remark}[Versatility of the results with respect to \player{C}'s objective]
The problem faced by \player{S} is a linear function of the covariance matrices $\{H_k\}$ since optimal reaction of \player{C} turns out to be linear in the posterior estimate of the control-free state, i.e., $\E\{\rx_k^o | \rs_{1:k}\}$, as seen in \eqref{eq:optU}. Therefore the results henceforth hold for any other scenarios where \player{C} has any other objective in which optimal reaction of \player{C} still turns out to be a linear function of the posterior estimate. Note that we need to compute the associated $\{\{\Xi^{\omega}_k\},\xi^{\omega}\}$ accordingly.\hfill $\square$
\end{remark}

\begin{remark}[Versatility of the results with respect to \player{S}'s objective]
We motivate and consider the case where \player{S} has objective \eqref{eq:Sobj}. However, the results henceforth would also hold for scenarios where \player{S}'s objective is any other (convex or non-convex) quadratic function of the state and control input. Note that we also need to compute the associated $\{\{\Xi^{\omega}_k\},\xi^{\omega}\}$ accordingly. \hfill $\square$
\end{remark}

Compared to original form of the optimization problem \eqref{eq:etaStar}, the optimization problem \eqref{eq:short} has a concrete structure showing that the optimization function depends on the signaling strategy through the covariance matrix only, and it is a linear function of the covariance matrix. Therefore it is instructive to study the relationship between the covariance matrix and the signaling strategy. Since the underlying distributions are all jointly Gaussian, we can express the covariance matrix in closed form:
\begin{equation}\label{eq:covariance}
H_k = \E\{\rx_k^o \rs_{1:k}'\}\E\{\rs_{1:k}\rs_{1:k}'\}^{\dagger}\E\{\rx_k^o \rs_{1:k}'\}'
\end{equation}
and the signal $\rs_k$ is given by \eqref{eq:signal}. This yields that even though computing robust signaling strategies would mean finding a certain number of matrices, it is a highly nonlinear and non-convex optimization problem due to the matrix inversion in \eqref{eq:covariance}. Therefore it is difficult to obtain a global solution through a generic attempt, e.g., genetic algorithm \cite{ref:Sadeghi14} or particle swarm optimization \cite{ref:Kennedy95}. On the other hand, the following proposition shows that there is a computationally tractable relationship between the covariance matrix and linear plus noise signaling strategies.

\begin{proposition}[A Necessary and Sufficient Condition]\label{prop:suff}
For any signaling rule $\eta\in\Eta$, covariance matrix of posterior estimate of the control-free state, $\{H_k = \cov\{\E\{\rx_{k}^o|\rs_{1:k}\}\}\}$, satisfies
\begin{subequations}\label{eq:neces}
\begin{align}
&\Sigma_1^o \succeq H_1 \succeq O,\\
&\Sigma_k^o \succeq H_k \succeq AH_{k-1}A',\mbox{ for } k>1,
\end{align}
\end{subequations}
where\footnote{By the definition of $\Sigma_k^o$, we have $\Sigma_k^o = A\Sigma_k^oA' + \Sigma_w$.} $\Sigma_k^o := \E\{\rx_k^o(\rx_k^o)'\}$.

Furthermore for any collection of positive semi-definite matrices $S_{1:\kappa}$ satisfying
\begin{subequations}\label{eq:constraint}
\begin{align}
&\Sigma_1^o \succeq S_1 \succeq O,\\
&\Sigma_k^o \succeq S_k \succeq AS_{k-1}A',\mbox{ for } k>1,
\end{align}
\end{subequations}
there exists a (memoryless) linear-plus-noise signaling strategy\footnote{Such a signaling strategy is described in \eqref{eq:optSignal} later.} $\eta\in\Eta$ such that $S_k = H_k$. 
\end{proposition}

In the following, we provide a description of signaling strategies\footnote{A derivation of the associated signaling strategies can be found in the constructive proof of Proposition \ref{prop:suff}, which is provided in Appendix \ref{app:suff}.} that lead to covariance matrices matching with a given collection of positive semi-definite matrices $S_{1:\kappa}$ satisfying \eqref{eq:constraint}. To this end, we let 
\begin{equation*}
\Sigma_k^o -AS_{k-1}A' = \bU_k\begin{bmatrix} \bLambda_k & O \\ O & O\end{bmatrix}\bU_k' 
\end{equation*}
be the eigen-decomposition such that $\bLambda_k \in\universe{S}_{++}^{t_k}$, where $t_k = \rank\{\Sigma_k^o -AS_{k-1}A'\}$. Furthermore, we let
\begin{align*}
T_k:= \begin{bmatrix} \bLambda_k^{1/2}& O \end{bmatrix} \bU_k' (S_k-AS_{k-1}A')\bU_k \begin{bmatrix} \bLambda_k^{1/2} \\ O \end{bmatrix}
\end{align*}
have the eigen-decomposition $T_k = U_k\Lambda_kU_k'$ with eigenvalues, e.g., $\lambda_{k,1},\ldots,\lambda_{k,t_k}$. We note that $\lambda_{k,t}$ turns out to be in $[0,1]$. Then, the associated signaling strategy is given by
\begin{equation}\label{eq:optSignal}
\rs_k = L_{k,k}' \rx_k + \rn_k,
\end{equation}
where $\rn_k\sim \Gauss(0,\Theta_k)$ with $\Theta_k = \diag\{\theta_{k,1}^2,\ldots,\theta_{k,t_k}^2,0,\ldots,0\}$, and the gain matrix $L_{k,k}$ is given by
\begin{equation*}
L_{k,k} = \bU_k\begin{bmatrix} \bLambda_k^{-1/2} U_k\Lambda_k^o & O \\ O & O \end{bmatrix},
\end{equation*}
where $\Lambda_k^o := \mathrm{diag}\{\lambda_{1,1}^o,\ldots,\lambda_{1,t_k}^o\}$ is a diagonal matrix such that $\{\lambda_{k,i}^o,\theta_{k,i}^2\}_{i=1}^{t_k}$ satisfies
\begin{equation*}
\frac{(\lambda_{k,i}^o)^2}{(\lambda_{k,i}^o)^2 + \theta_{k,i}^2} = \lambda_{k,i}, \forall\, i=1.\ldots,t_k.
\end{equation*}

The following corollary to Proposition \ref{prop:suff} provides a problem equivalent to \eqref{eq:short}.

\begin{corollary}[Equivalence Result]\label{cor:equivalent}
The problem faced by \player{S}, i.e., \eqref{eq:Sobj}, is equivalent to
\begin{equation}\label{eq:main}
\min_{\{S_k\in\universe{S}_+^m\}} \max_{p\in\Delta(\bOmega)}  \sum_{\omega\in\bOmega}p_{\omega} \left(\xi^{\omega} + \sum_{k=1}^{\kappa}\trace\{S_k \Xi_k^{\omega}\}\right),
\end{equation}
subject to the following linear matrix inequalities:
\begin{subequations}\label{eq:psi}
\begin{align}
&\Sigma_1^o \succeq S_1 \succeq O\\
&\Sigma_k^o \succeq S_k \succeq AS_{k-1}A',\mbox{ for } k>1.
\end{align}
\end{subequations}
And given a solution $S_{1:\kappa}$, an associated signaling strategy can be computed according to \eqref{eq:optSignal}.
\end{corollary}

Henceforth, we will be working with \eqref{eq:main} instead of \eqref{eq:short} while analyzing the equilibrium of the game $\calG$. Furthermore, for brevity of presentation, let us introduce
\begin{equation*}
S := \begin{bmatrix} S_{\kappa} & & \\ & \ddots & \\ & & S_1 \end{bmatrix},\; \Xi^{\omega} := \begin{bmatrix} \Xi_{\kappa}^{\omega} & & \\ & \ddots & \\ & & \Xi_1^{\omega} \end{bmatrix},
\end{equation*}
and $\Psi \subset \universe{S}_+^{m\kappa}$ be the set corresponding to the constraints \eqref{eq:psi} in this new high-dimensional space, i.e., $\R^{m\kappa \times m\kappa}$. With this new notation, the problem faced by \player{S}, i.e., \eqref{eq:main}, can be written as
\begin{align}\label{eq:maxmin}
\min_{S\in\Psi}\max_{p\in\Delta(\bOmega)}\;\sum_{\omega\in\bOmega}p_{\omega}\left(\xi^{\omega}+\trace\left\{S\Xi^{\omega}\right\}\right).
\end{align}

The following proposition addresses the existence of an equilibrium for the game $\calG$.

\begin{proposition}[Existence Result]\label{prop:existence}
There exists at least one tuple of strategies $(\eta,\{\gamma^{\omega}(\eta)\}_{\omega\in\bOmega})$ attaining the equilibrium of the Stackelberg game $\calG$, i.e., satisfying \eqref{eq:equilibrium}. 
\end{proposition}

It is instructive to examine whether optimal signaling strategies end up using irrelevant information or not. The inner optimization problem in \eqref{eq:maxmin}
\begin{equation}\label{eq:inner}
\max_{p\in\Delta(\bOmega)}\sum_{\omega\in\bOmega} p_{\omega} \left(\xi^{\omega} + \trace\{S\Xi^{\omega}\}\right)
\end{equation}
is a convex function of $S\in\Psi$ since the maximum of any family of linear/affine functions is a convex function \cite{ref:Boyd04}. Therefore, there might be examples where any extreme point of the constraint set $\Psi$ is not a solution for \eqref{eq:maxmin}. Correspondingly, optimal signaling strategy can turn out to be including irrelevant information. This is interesting in view of Blackwell's Irrelevant Information Theorem \cite[Theorem D.1.1]{ref:Yuksel13}. Particularly, the theorem says that given a cost measure, for any Borel measurable function that uses irrelevant information, there exists another Borel measurable function that does not use any irrelevant information and can lead to a cost less than or equal to the one attained with the former function. Therefore we can conclude that in this robust setting, linear signaling strategies are not the best one within the general class of measurable strategies.

Next, we seek to compute the equilibrium of $\calG$. To this end, we examine the equilibrium conditions further. In particular, according to \eqref{eq:maxmin}, given $S\in\Psi$, maximizing $p^*\in\Delta(\bOmega)$ is given by
\begin{align}
p^{*} \in \bigg\{p&\in\Delta(\bOmega)\,|\,p_{\omega} = 0 \mbox{ provided that} \nn\\
&\left(\xi^{\omega} + \trace\{S\Xi^{\omega}\}\right)<\max_{\omega'}\left(\xi^{\omega'} + \trace\{S\Xi^{\omega'}\}\right)\bigg\}\label{eq:postar}
\end{align}
since the optimization objective in \eqref{eq:maxmin} is a linear function of $p\in\Delta(\bOmega)$. Based on the observation \eqref{eq:postar} and the assumption that the type set $\bOmega$ is finite, the following theorem provides an algorithm to compute robust sensor outputs.

\begin{theorem}[Computing the Equilibrium]\label{theorem:compute}
The value of the Stackelberg equilibrium \eqref{eq:maxmin} is given by $\mu = \min_{\omega\in\bOmega}\{\mu_{\omega}\}$, where 
\begin{align}\label{eq:vj}
\mu_{\omega} := \min_{S\in\Psi}&\; \left[\xi^{\omega} + \trace\big\{\Xi^{\omega}S\big\}\right]\\ 
\mathrm{s.t. }&  \;\trace\big\{S(\Xi^{\omega}-\Xi^{\omega'})\big\} \geq \xi^{\omega'}-\xi^{\omega}\; \forall \omega'\in\bOmega.\nn
\end{align}

Furthermore, let $\mu_{\omega^*} = \mu$ and
\begin{align}
S^*\in \argmin_{S\in\Psi}&\; \left[\xi^{\omega} + \trace\big\{\Xi^{\omega}S\big\}\right]\label{eq:Sstarj}\\ 
\mathrm{s.t. }&  \;\trace\big\{S(\Xi^{\omega}-\Xi^{\omega'})\big\} \geq \xi^{\omega'}-\xi^{\omega}\; \forall \omega'\in\bOmega.\nn
\end{align}
Then, given $S^*\in\Psi$, an associated linear-plus-noise signaling strategy $\eta\in\Eta$ can be computed according to \eqref{eq:optSignal}.
\end{theorem}

For the reader's reference, in the following, we list the steps to compute robust sensor design strategies:
\begin{itemize}
\item We first compute the matrices $\Xi_k^{\omega}$, described in \eqref{eq:V}, and scalars $\xi^{\omega}$, described in \eqref{eq:xi1} and \eqref{eq:xi2}, for each type of control objectives, $\omega\in\bOmega$. This step includes computation of gain matrices of optimal control input, e.g., \eqref{eq:optU}, in an LQG control problem. 
\item Given computed $\{\{\Xi_k^{\omega},\xi^{\omega}\}\}$, we solve SDP, described in \eqref{eq:vj}, for each type by using an SDP solver, e.g., \cite{ref:CVX,ref:Grant08}, numerically. 
\item When we compute $S^*$ according to \eqref{eq:Sstarj}, we can compute the associated signaling strategies according to (30). Note that this step includes computation of eigen-decomposition of some matrices.
\end{itemize}
We re-emphasized that we provide an algorithm to compute robust signaling strategies {\bf globally} even though the problem is highly nonlinear and non-convex. There is, however, still room to develop computationally more efficient approaches. And the solution concept proposed in the paper can be used as a benchmark to evaluate performance of such computationally efficient algorithms. 

\section{Noisy or Partial Measurements}\label{sec:noisy}

In this section, we seek to obtain robust signaling strategies under noisy or partial (i.e., imperfect) measurements of the type \eqref{eq:measurement}. To this end, we turn the problem to the same structure with the case under perfect measurements based on a recent result from \cite{ref:Sayin18d} and then invoke the results from the previous section. 

There are several challenges in robust sensor design under imperfect measurements. For example, Proposition \ref{prop:suff} does not hold anymore. To obtain a result similar to Proposition \ref{prop:suff}, a first attempt would be to focus on the covariance matrix of the posterior estimate of control-free {\em measurements}, i.e., $\cov\{\E\{\ry_k^o|\rs_{1:k}\}\}$, where $\ry_k^o := C\rx_k^o + \rv_k$, instead of $H_k = \cov\{\E\{\rx_k^o|\rs_{1:k}\}\}$. Quite contrary to the control-free state $\{\rx_k^o\}$, however, control-free measurements $\{\ry_k^o\}$ do not necessarily constitute a Markov process since in general $\E\{\ry_{k}^o|\ry_{1:k-1}^o\}\neq \E\{\ry_k^o|\ry_{k-1}^o\}$. Therefore, we focus on the augmented vector of measurements $\ry_{1:k}^o\in\R^{nk}$. 

Since the measurements are jointly Gaussian, we have
\begin{align}\nn
\E\{\ry_k^o|\ry_{1:k-1}^o\} = \E\{\ry_k^o(\ry_{1:k-1}^o)'\}\E\{\ry_{1:k-1}^o(\ry_{1:k-1}^o)'\}^{\dagger} \ry_{1:k-1}^o.
\end{align}
This implies that the augmented measurements $\{\ry_{1:k}\}_{k\geq 0}$ evolve according to
\begin{equation}
\ry_{1:k}^o = A_k \ry_{1:k-1}^o + \re_k, 
\end{equation}
where we define
\begin{align*}
&A_k := \begin{bmatrix} \E\{\ry_k^o(\ry_{1:k-1}^o)'\}\E\{\ry_{1:k-1}^o(\ry_{1:k-1}^o)'\}^{\dagger} \\ I \end{bmatrix},\\
&\re_k := \begin{bmatrix} \ry_k^o - \E\{\ry_k^o|\ry_{1:k-1}^o\} \\ \mathbf{0} \\ \end{bmatrix}
\end{align*}
Note that neither $A_k\in\R^{nk\times n(k-1)}$ nor $\re_k\in\R^{nk}$ depend on signaling or control strategies. Furthermore, similar to \eqref{eq:so}, it can be shown that
\begin{equation}\label{eq:soy}
\E\{\ry_{1:k}^o | \rs_{1:k}\} = \E\{\ry_{1:k}^o | \rs_{1:k}^o\},
\end{equation}
where we now have $\rs_k^o = L_{k}'\ry_{1:k}^o + \rn_k$.

Since $\rs_{1:k}^o$ is $\sigma$-$\ry_{1:k}$ measurable, $\rx_k^o$ and $\rs_{1:k}^o$ are independent of each other conditioned on $\ry_{1:k}^o$. This implies that $\rx_k^o$, $\ry_{1:k}^o$, and $\rs_{1:k}^o$ can be viewed as forming a Markov chain in the order $\rx_k^o \rightarrow \ry_{1:k}^o \rightarrow \rs_{1:k}^o$. In that respect, \cite[Lemma 1]{ref:Sayin18d} shows that when $\{\rx_k^o,\ry_{1:k}^o,\rs_{1:k}\}$ are jointly Gaussian and form a Markov chain in the order $\rx_k^o \rightarrow \ry_{1:k}^o \rightarrow \rs_{1:k}^o$,
there exists a linear relation between $\E\{\rx_k^o|\rs_{1:k}^o\}$ and $\E\{\ry_{1:k}^o|\rs_{1:k}^o\}$ irrespective of $\rs_{1:k}^o$, and the relation is given by
\begin{equation}\label{eq:condexp2}
\E\{\rx_k^o|\rs_{1:k}^o\} = D_k\E\{\ry_{1:k}^o|\rs_{1:k}^o\},
\end{equation}
where $D_k\in\R^{m\times nk}$ is defined by
\begin{equation*}
D_k := \E\{\rx_k^o(\ry_{1:k}^o)'\}\E\{\ry_{1:k}^o(\ry_{1:k}^o)'\}^{\dagger}.
\end{equation*}

Under imperfect measurements, counterpart of the covariance matrix $H_k = \cov\{\E\{\rx_k^o|\rs_{1:k}\}\}$  is the covariance matrix of the posterior estimate of control-free augmented measurements, denoted by
\begin{equation*}
Y_k := \cov\{\E\{\ry_{1:k}^o | \rs_{1:k}\}\}.
\end{equation*}
Based on \eqref{eq:soy} and \eqref{eq:condexp2}, we have
\begin{equation*}
H_k = D_kY_kD_k'. 
\end{equation*}
Correspondingly, \eqref{eq:short}, i.e., the problem faced by \player{S}, can be written as
\begin{equation}\label{eq:shorty}
\min_{\eta\in\Eta}\max_{p\in\Delta(\bOmega)} \sum_{\omega\in\bOmega} p_{\omega} \left(\xi^{\omega} + \sum_{k=1}^{\kappa} \trace\left\{Y_k W_k^{\omega}\right\}\right),
\end{equation}
where we define\footnote{We use the cyclic property of the trace operator.} $W_k^{\omega} := D_k'\Xi^{\omega}D_k$, which can be viewed as the counterpart of $\Xi_k^{\omega}$ under imperfect measurements.  We remark the resemblance between \eqref{eq:short} and \eqref{eq:shorty}, where we have $Y_k$ instead of $H_k$ and $W_k^{\omega}$ instead of $\Xi_k^{\omega}$. 

Recall that $\ry_k \in \R^n$, which yields that $\ry_{1:k} \in \R^{nk}$. Under the assumption that $\rs_k \in \R^{nk}$ instead of $\rs_k\in\R^m$ (so that \player{S} can disclose the auxiliary $\ry_{1:k}$ perfectly), we would have transformed the problem under imperfect measurements into a problem under perfect measurements, and correspondingly we could have invoked the results from the previous section directly. The following lemma shows that results for the case where $\rs_k\in\R^{nk}$ would also hold for the case where $\rs_k\in\R^m$ even when $nk>m$.

\begin{lemma}\label{lem:measurements}
Let us denote signaling strategies when $\rs_k \in \R^{nk}$ by $\teta_k$ and the associated strategy space by $\Teta_k$. Then for any $\{\teta_k\in\Teta_k\}$, there exists a $\{\eta_k\in\Eta_k\}$ such that they both lead to the same control strategy and correspondingly the same cost for \player{S}. And such a signaling strategy $\{\eta_k\in\Eta_k\}$ is given by
\begin{equation}\label{eq:teta}
\eta_k(\ry_{1:k}) = \E\{\rx_k^o|\teta_1(\ry_1),\ldots,\teta_k(\ry_{1:k})\},\;\forall k\geq 1.
\end{equation} 
\end{lemma}

Based on Lemma \ref{lem:measurements}, the following corollary to Proposition \ref{prop:suff} provides a tractable necessary and sufficient condition on $\{Y_k\}$.

\begin{corollary}[A Necessary and Sufficient Condition Under Imperfect Measurements]\label{cor:lem}
For any signaling rule $\eta\in\Eta$, covariance matrix of posterior estimate of the control-free measurements, $\{Y_k = \cov\{\E\{\ry_{1:k}^o|\rs_{1:k}\}\}\}$, satisfies
\begin{subequations}\label{eq:neces}
\begin{align}
&\Sigma_1^y \succeq Y_1 \succeq O,\\
&\Sigma_k^y \succeq Y_k \succeq A_{k-1}Y_{k-1}A_{k-1}',\mbox{ for } k>1,
\end{align}
\end{subequations}
where $\Sigma_k^y := \E\{\ry_{1:k}^o(\ry_{1:k}^o)'\}$.

Furthermore for any collection of positive semi-definite matrices $S_{1:\kappa}$ satisfying
\begin{subequations}\label{eq:constraint}
\begin{align}
&\Sigma_1^y \succeq S_1 \succeq O,\\
&\Sigma_k^y \succeq S_k \succeq A_{k-1}S_{k-1}A_{k-1}',\mbox{ for } k>1,
\end{align}
\end{subequations}
there exists a linear-plus-noise signaling strategy $\eta\in\Eta$ such that $S_k = Y_k$. 
\end{corollary}

The following corollary provides a problem equivalent to \eqref{eq:shorty}.

\begin{corollary}\label{cor:equi}
The problem faced by \player{S}, i.e., \eqref{eq:shorty}, is equivalent to  
\begin{equation}\label{eq:SDPimp}
\min_{\{S_k\in\universe{S}^{kn}_+\}}\max_{p\in\Delta(\bOmega)} \sum_{\omega\in\bOmega} p_{\omega} \left(\xi^{\omega} + \sum_{k=1}^{\kappa} \trace\left\{S_k W_k^{\omega}\right\}\right),
\end{equation}
subject to the following linear matrix inequalities:
\begin{subequations}\label{eq:cons2}
\begin{align}
&\Sigma_1^y \succeq S_1 \succeq O,\\
&\Sigma_k^y \succeq S_k \succeq A_{k-1}S_{k-1}A_{k-1}', \mbox{ for } k>1.
\end{align}
\end{subequations}

Furthermore, given as solution $\{S_{1:\kappa}\}$ for \eqref{eq:SDPimp}, we can compute an optimal signaling strategy for \eqref{eq:shorty}, as follows:
\begin{itemize}
\item Compute signaling strategies $\teta\in\Teta$ as if signal $\rs_k$ can be $nk$ dimensional, i.e., $\rs_k\in\R^{nk}$, according to \eqref{eq:optSignal}, where we have $\Sigma_k^y$ instead of $\Sigma_k^o$, $A_k$ instead of $A$, and $\ry_{1:k}$ instead of $\rx_k$.\footnote{We provide closed-form expressions for the auxiliary parameters $\Sigma_k^y, A_k$, and $D_k$ in Appendix \ref{app:auxiliary} for the reader's reference.}
\item For computed $\teta\in\Teta$, compute associated signaling strategies $\eta\in\Eta$ according to \eqref{eq:teta}.
\end{itemize}
\end{corollary}

We remark that under imperfect measurements optimal signaling strategies are not necessarily memoryless anymore. 

Through a similar notational convention as in the previous section, we can write \eqref{eq:SDPimp} as
\begin{equation}\label{eq:maxmin2}
\min_{S\in\bar{\Psi}} \max_{p\in\bOmega} \sum_{\omega\in\bOmega} p_{\omega} \left(\xi^{\omega} + \trace\left\{S W^{\omega}\right\}\right),
\end{equation}
where we let $\bar{\Psi}\subset \universe{S}^{m\kappa(\kappa+1)/2}$ be the set corresponding to the constraints \eqref{eq:cons2}. Then, based on Corollary \ref{cor:equi}, the following corollary to Theorem \ref{theorem:compute} provides a computationally tractable way to compute robust sensor outputs under imperfect measurements.

\begin{corollary}[Computing the Equilibrium Under Imperfect Measurements] \label{cor:compute}

The value of the Stackelberg equilibrium \eqref{eq:maxmin2} is given by $\mu = \min_{\omega\in\bOmega}\{\mu_{\omega}\}$, where 
\begin{align}
\mu_{\omega} := \min_{S\in\bar{\Psi}}&\; \left[\xi^{\omega} + \trace\big\{W^{\omega}S\big\}\right]\\ 
\mathrm{s.t. }&  \;\trace\big\{(W^{\omega}-W^{\omega'})S\big\} \geq \xi^{\omega'} - \xi^{\omega}\; \forall \omega'\in\bOmega.\nn
\end{align}

Furthermore, let $\mu_{\omega^*} = \mu$ and
\begin{align}
S^*\in \argmin_{S\in\bar{\Psi}}&\; \left[\xi^{\omega} + \trace\big\{W^{\omega}S\big\}\right]\\ 
\mathrm{s.t. }&  \;\trace\big\{(W^{\omega}-W^{\omega'})S\big\} \geq \xi^{\omega'} - \xi^{\omega}\; \forall \omega'\in\bOmega.\nn
\end{align}
Then, given $S^*\in\bPsi$, an associated linear-plus-noise signaling strategy $\eta\in\Eta$ can be computed according to Corollary \ref{cor:equi}.
\end{corollary}

\section{Illustrative Examples}\label{sec:examples}

In this section, we examine the performance of the proposed defense measure over various attack scenarios. As an illustrative example, we set length of the time horizon at $\kappa=10$, dimension of state $m=2$, and dimension of control input $r=2$. We consider the scenario where the system's control objective is to track an exogenous process $\{\rz_k\}$ evolving according to
\begin{equation*}
\rz_{k+1} = A^z\rz_k + \rw_k^z,
\end{equation*}
where $\rz_1\sim\Gauss(0,I_2)$, and $\{\rw_k^z\sim\Gauss(0,I_2)\}$ is a white Gaussian noise process, and they are independent of each other and every other parameter. Correspondingly, the system's control objective can be written as
\begin{equation}\label{eq:Cnew}
U_C^{\omega_o}(\eta,\gamma^{\omega_o}) = \sum_{k=1}^{\kappa} \E\|\rx_{k+1}^{\omega_o}-\rz_{k+1}\|^2 + \E\|\ru_k^{\omega_o}\|^2.
\end{equation}
We set $A = A^z = I_2$ and $B= \begin{bmatrix} 1&1\\ 0 & 1\end{bmatrix}$ while $\Sigma_1 = I_2$ and $\Sigma_w = I_2$. The sensor has access to the measurements:
\begin{align*}
&\ry_k = C\rx_k + \rv_k\\
&\ry_k^z = C^z \rz_k + \rv_k^z,
\end{align*}
where $\{\rv_k^z\}$ is a white Gaussian measurement noise independent of every other
parameter. 

Note that the control objective is not in the form of \eqref{eq:Cobj}; however, we can transform it into the form of \eqref{eq:Cobj} by introducing the augmented state $\rtx_k :=\begin{bmatrix} \rx_k' & \rz_k' \end{bmatrix}'$ evolving according to
\begin{equation*}
\begin{bmatrix} \rx_{k+1} \\ \rz_{k+1} \end{bmatrix} = \begin{bmatrix} I & O \\ O & I\end{bmatrix}\begin{bmatrix} \rx_k \\ \rz_k \end{bmatrix} + \begin{bmatrix} I \\ O \end{bmatrix}\ru_k + \begin{bmatrix} \rw_k \\ \rw_k^z\end{bmatrix}
\end{equation*}
and augmented measurements are then given by
\begin{equation*}
\rty_k = \begin{bmatrix} C & O \\ O & C_z \end{bmatrix}\begin{bmatrix} \rx_k \\ \rz_k \end{bmatrix} + \begin{bmatrix} \rv_k \\ \rv_k^z \end{bmatrix}.
\end{equation*}
Correspondingly \eqref{eq:Cnew} can be written as
\begin{equation*}
\sum_{k=1}^{\kappa} \E\|\rtx_{k+1}\|_{Q}^2 + \E\|\ru_k\|^2,
\end{equation*}
where
\begin{equation*}
Q := \begin{bmatrix} I_2 \\ -I_2\end{bmatrix}\begin{bmatrix} I_2 & -I_2\end{bmatrix}.
\end{equation*}

As examples of attack scenarios, we consider a type set $\Omega = \{\omega_a,\omega_b,\omega_c\}$. Let us partition the underlying state $\rx_k = \begin{bmatrix} \rx_{k,1} & \rx_{k,2}\end{bmatrix}'$ and the exogenous process $\rz_k = \begin{bmatrix} \rz_{k,1}&\rz_{k,2}\end{bmatrix}'$, where $\rx_{k,1}$ and $\rz_{k,1}$ (or $\rx_{k,2}$ and $\rz_{k,2}$) correspond to the first (or the second) entries of the state and the exogenous process, respectively. We assume that type-$\omega_{a}$ attacker seeks to make $\{\rx_{k,1}\}$ track $\{-\rz_{k,1}\}$ instead of $\{\rz_{k,1}\}$ whereas it is not interested in $\{\rx_{k,2}\}$. Then its control objective can be written as
\begin{equation*}
U_C^{\omega_a}(\eta,\gamma^{\omega_a}) = \sum_{k=1}^{\kappa} \E\|\rx_{k+1,1}^{\omega_{a}}-(-\rz_{k+1,1})\|^2 + \E\|\ru_k^{\omega_{a}}\|^2.
\end{equation*}
Similarly type-$\omega_{b}$ attacker seeks to make $\{\rx_{k,2}\}$ track $\{-\rz_{k,2}\}$ instead of $\{\rz_{k,2}\}$ whereas it is not interested in $\{\rx_{k,1}\}$. Then its control objective can be written as
\begin{equation*}
U_C^{\omega_b}(\eta,\gamma^{\omega_b}) = \sum_{k=1}^{\kappa} \E\|\rx_{k+1,2}^{\omega_{b}}-(-\rz_{k+1,2})\|^2 + \E\|\ru_k^{\omega_{b}}\|^2.
\end{equation*}
On the other hand, type-$\omega_c$ attacker is interested in both entries and seeks to make the entire state $\{\rx_k\}$ track $\{-\rz_k\}$. Accordingly, its control objective can be written as
\begin{equation*}
U_C^{\omega_c}(\eta,\gamma^{\omega_c}) = \sum_{k=1}^{\kappa} \E\|\rx_{k+1}^{\omega_{c}}-(-\rz_{k+1})\|^2 + \E\|\ru_k^{\omega_{c}}\|^2.
\end{equation*}
We note that numerical simulations show that mixtures of types $\omega_a$, $\omega_b$ and $\omega_c$ can lead to larger costs for the system than any single type, including type-$\omega_c$. In the following, we examine the cost of \player{S} under perfect and imperfect measurements separately. 

\begin{table}[t!]
\renewcommand{\arraystretch}{1.4}
\caption{Under perfect measurements, i.e., $\rty_k = \rtx_k$, comparison of the costs to \player{S}, i.e., $U_S$ as described in \eqref{eq:Sobj}, over various scenarios. Entries at each row corresponds to the cost to \player{S} for different types of \player{C} when it constructs the sensor outputs according to the extended type set $\bOmega$. And the last column provides the maximum cost of \player{S} across all types of \player{C}.}\label{tab:perfect}
\begin{center}
\begin{tabular}{l||c||c|c|c||c}
$\bOmega$ & $\omega_o$ & $\omega_{a}$ & $\omega_{b}$ & $\omega_c$ & Max \\
\hline
$\{\omega_o\}$ & \colorcell $43.03$ & $323.70$ & $\mathbf{354.02}$ & $352.65$ & $\mathbf{354.02}$ \\
\hline
$\{\omega_a\}$ & $122.31$ & \colorcell $191.12$ & $\mathbf{351.85}$ & $246.30$ & $\mathbf{351.85}$\\
$\{\omega_b\}$ & $119.09$ & $\mathbf{320.81}$ & \colorcell $185.63$ & $267.35$ & $\mathbf{320.81}$\\
$\{\omega_c\}$ & $106.26$ & $202.40$ & $\mathbf{224.77}$ & \colorcell $137.71$ & $\mathbf{224.77}$\\
\hline
$\{\omega_o,\omega_a\}$ & \colorcell $119.65$ & \colorcell $191.12$ & $\mathbf{351.11}$ & $245.95$ & $\mathbf{351.11}$\\
$\{\omega_o,\omega_b\}$ & \colorcell $115.29$ & $\mathbf{321.56}$ & \colorcell $185.63$ & $267.93$ & $\mathbf{321.56}$\\
$\{\omega_o,\omega_c\}$ & \colorcell $106.26$ & $202.40$ & $\mathbf{224.77}$ & \colorcell $137.71$ & $\mathbf{224.77}$\\
\hline
$\{\omega_o,\omega_a,\omega_b,\omega_c\}$ & \colorcell $81.17$ & \colorcell $\mathbf{199.46}$ & \colorcell $\mathbf{199.46}$ & \colorcell $166.68$ & $\mathbf{199.46}$
\end{tabular}
\end{center}
\end{table}

Under perfect measurements, the sensor has access to $\rty_k = \rtx_k$. Note that perfect measurements provide the utmost {\em freedom} for \player{S} to shape the belief of the attacker. For any cost that \player{S} can attain under imperfect measurements, \player{S} can select a signaling strategy under perfect measurements to attain the same cost. Therefore, \player{S} attains the lowest possible cost under perfect measurements in the robust sensor design framework. 

In Table \ref{tab:perfect}, we tabulate the cost to \player{S} for the scenarios where $i)$ there is no attack, i.e., type of \player{C} is $\omega_o$; $ii)$ there is an attack by type-$\omega_a$ attacker, i.e., type of \player{C} is $\omega_a$; $iii)$ there is an attack by type-$\omega_b$ attacker, i.e., type of \player{C} is $\omega_b$; and $iv)$ there is an attack by type-$\omega_c$ attacker, i.e., type of \player{C} is $\omega_c$. Cost to \player{S} varies depending on the type of \player{C} and how prepared \player{S} is while constructing the sensor outputs. In other words, \player{S} constructs the sensor outputs according to an extended type set. 

For example, \player{S} would have constructed signaling strategies according to $\bOmega = \{\omega_o\}$ if it views that there would not be any attack. Correspondingly, if there is no attack, then the cost would be $43.03$. However, if there is an attack by, e.g., type-$\omega_b$ attacker, then the cost would be $354.02$, which is significantly higher compared to $43.03$. Next, consider that \player{S} has constructed the sensor outputs according to $\bOmega = \{\omega_b\}$. Then, the system would be prepared against an attack by type-$\omega_b$ attacker, and the cost would be $185.63$ when type-$\omega_b$ attacker attacks. This is significantly lower than the cost $354.02$ obtained when \player{S} constructs sensor outputs without any concern about possible attacks. However, now the cost to \player{S} is $119.09$ if there is no attack and the type of \player{C} is $\omega_o$. This is also higher than the cost $43.3$ obtained when \player{S} constructs sensor outputs without any concern about possible attacks. It is an uncertainty whether there will be an attack or not. Correspondingly, if \player{S} constructs the sensor outputs according to $\bOmega = \{\omega_o,\omega_b\}$, then the cost would be $115.29$ when there is no attack. It is lower than the cost $119.09$ obtained before. On the other hand, the cost would still be around $185.63$ when type-$\omega_b$ attacker attacks the system. 

Even though \player{S} is prepared to an attack by type-$\omega_b$ attacker by constructing the sensor outputs according to $\bOmega = \{\omega_o,\omega_b\}$, there can be an attack by another type attacker, e.g., type-$\omega_a$. Then the cost would be $321.56$, which is significantly higher than the cost $185.63$ that would be obtained when type-$\omega_b$ attacker attacks. The system can decrease this cost by also considering the possibility of attacks by type-$\omega_a$ attacker while constructing the sensor outputs. For example, if \player{S} constructs the sensor outputs by taking into account types $\omega_a$, $\omega_b$, and $\omega_c$ attacks, then the cost would be at most $199.46$ if any of those types of attacks occurs and the cost would be $81.17$ if there is no attack. These examples show the importance of constructing sensor outputs in a robust way.

\begin{table}[t!]
\renewcommand{\arraystretch}{1.4}
\caption{Under imperfect measurements, comparison of the costs to \player{S}, i.e., $U_S$ as described in \eqref{eq:Sobj}, over various scenarios. Entries at each row corresponds to the cost to \player{S} for different types of \player{C} when it constructs the sensor outputs according to the extended type set $\bOmega$. And the last column provides the maximum cost of \player{S} across all types of \player{C}.}\label{tab:imperfect}
\begin{center}
\begin{tabular}{l||c||c|c|c||c}
$\bOmega$ & $\omega_o$ & $\omega_{a}$ & $\omega_{b}$ & $\omega_c$ & Max \\
\hline
$\{\omega_o\}$ & \colorcell  $112.39$ & $326.21$ & $324.82$ & $\mathbf{403.09}$ & $\mathbf{403.09}$ \\
\hline
$\{\omega_a\}$ & $171.64$ &\colorcell $209.99$ & $\mathbf{259.51}$ & $242.26$ & $\mathbf{259.56}$\\
$\{\omega_b\}$ & $167.80$ & $\mathbf{269.60}$ & \colorcell $195.90$ & $254.20$ & $\mathbf{269.60}$\\
$\{\omega_c\}$ & $187.26$ & $\mathbf{211.62}$ & $201.94$ & \colorcell $199.51$ & $\mathbf{211.62}$\\
\hline
$\{\omega_o,\omega_a\}$ & \colorcell $171.40$ & \colorcell $209.99$ & $\mathbf{260.40}$ & $242.97$ & $\mathbf{260.40}$\\
$\{\omega_o,\omega_b\}$ & \colorcell $167.53$ & $\mathbf{270.40}$ & \colorcell $195.90$ & $255.04$ & $\mathbf{270.40}$\\
$\{\omega_o,\omega_c\}$ & \colorcell $187.26$ & $\mathbf{211.62}$ & $201.94$ & \colorcell $199.51$ & $\mathbf{211.62}$\\
\hline
$\{\omega_o,\omega_a,\omega_b,\omega_c\}$ & \colorcell $185.48$ & \colorcell $\mathbf{210.00}$ & \colorcell $\mathbf{210.00}$ & \colorcell $202.77$ &  $\mathbf{210.00}$
\end{tabular}
\end{center}
\end{table}

As an example for imperfect measurements, we take $\ry_k = \rx_{k,1} + \rx_{k,2} + \rv_k$, where $\rv_k\sim\Gauss(0,1)$, and $\ry_k^z = \rz_k + \rv_k^z$, where $\{\rv_k^z \sim\Gauss(0,I_2)\}$. In Table \ref{tab:imperfect}, we tabulate the costs to \player{S} over various scenarios, similar to Table \ref{tab:perfect}. Table \ref{tab:imperfect} shows that imperfect measurements lead to larger cost for the system when there is no attack, as to be expected. However, at certain scenarios, imperfect measurements can lead to better performance for the system. For example, when \player{S} constructs sensor outputs by considering that there would not be any attack, i.e., according to $\bOmega = \{\omega_o\}$, and there is an attack by type-$\omega_b$ attacker, the cost would be $324.82$, which is lower than the cost $354.02$ obtained under perfect measurements. For this scenario, imperfect measurements end up obfuscating type-$\omega_b$ attacker and lead to lower cost. In that respect, robust sensor outputs can be viewed as optimal imperfect measurements that lead to minimum cost for the system. Furthermore, the cost to \player{S} increases under imperfect measurements over the scenarios where it is prepared. In Table \ref{tab:imperfect}, we highlight those scenarios by shading their cells. A comparison of shaded cells of Tables \ref{tab:perfect} and \ref{tab:imperfect} verifies the observation emphasized above, i.e., imperfect measurements limit \player{S}'s ability to persuade \player{C}.

\section{Concluding remarks} \label{sec:conclusion}

In this paper, we have proposed and addressed persuasion-based robust sensor design as a security measure in control systems against attackers with unknown control objectives. By designing sensor outputs cautiously in advance, we have sought to shape attackers' believes about the underlying state of the system in order to induce them to act/attack to the system in line with the system's normal operation. We have modeled the problem formally under the solution concept of Stackelberg equilibrium where the defender/sensor is the leader. Non-strategic reaction of the follower/attacker implies that the defender faces an optimization problem while seeking to design robust sensor strategies. We have shown that the optimization problem is non-convex and highly nonlinear. To mitigate this issue, we have formulated a tractable problem equivalent to that problem and shown how to compute the associated signaling strategies. We have also extended the results to scenarios where there are imperfect measurements of the underlying state. Finally, we have examined the performance of the proposed framework across various attack scenarios.

Future directions of research on this topic include development of computationally efficient algorithms to compute optimal signaling rules and developing persuasion-based sensor design strategies for scenarios where attackers have partial information about the underlying state dynamics instead of full knowledge of it or have side information about the state instead of relying on sensor outputs only. Another interesting research direction would be its application on sensor placement or sensor selection. 

Furthermore, even though we have motivated the framework by relating it to security, the framework could also address strategic information disclosure over multi-agent control networks where agents have different control objectives. Particularly, independent of how we motivate and set up the signaling problem (e.g., a security application or a multi-agent non-cooperative control network), the solution concept developed can be adopted in various settings in a straightforward way provided that
\begin{itemize}
\item Information of interest and all random variables/vectors are jointly Gaussian
\item There is a single sender and possibly multiple receivers
\item Optimal reaction of each receiver is linear in its posterior belief
\item The sender's objective depends on receivers' reactions only through an arbitrary quadratic function of their posterior beliefs
\end{itemize}
In this paper, we have used this result to address uncertainties regarding attackers' (or receivers') objectives in the security of control systems over a finite horizon. However, the result could be adopted in several other scenarios as well, such as:
\begin{itemize}
\item over infinite horizon (as we did in \cite{ref:Sayin19d})
\item when there are additional tractable constraints on the covariance of the posterior belief (as we did in \cite{ref:Sayin18d} for a power constraint over the signals when there is an additive Gaussian noise channel between the sender and the receiver)
\end{itemize}
Furthermore, the ability to turn signaling problems (which lead to highly nonlinear and non-convex optimization problems) into linear optimization problems (over the space of positive-semi definite matrices with linear matrix inequality constraints) facilitates analysis of problems over more complex settings, e.g., where
\begin{itemize}
\item There can be multiple controllers seeking to drive the same system
\item There can be multiple senders that compete with each other to induce a controller to take certain actions
\end{itemize}


\appendices

\section{Novelty Relative to Reference \cite{ref:Sayin17b}}\label{app:comparison}

In Reference \cite{ref:Sayin17b}, we formulated an SDP equivalent to the original optimization problem in scenarios where there is no uncertainty on the attacker's objective. Although it may seem to have a similar flavor, in \cite{ref:Sayin17b} we used different technical tools and these tools cannot be adopted for the settings of this paper. Particularly, in \cite{ref:Sayin17b} we exploited the fact that a solution of a linear optimization problem over a compact convex set lies at extreme points\footnote{We say that a point in a convex set is an extreme point if it cannot be expressed as a convex combination of any other two points in the set.} of the constraint set. Even though we were able to characterize the extreme points of the constraint set for the specific optimization problem in \cite{ref:Sayin17b}, characterization of extreme points is challenging in general, e.g., see \cite{ref:Jerison54}. Furthermore, in the settings of this paper, the associated optimization problem includes an inner maximization induced by the sensor's robustness concern. Therefore the techniques developed in \cite{ref:Sayin17b} cannot be used in this setting since the objective function is no longer linear in the optimization arguments due to the inner maximization. It is indeed a convex function since maximum of any family of linear/affine functions is a convex function \cite{ref:Boyd04}. Correspondingly, the solution does not necessarily lie at an extreme point of the constraint set. To be able to solve this optimization problem globally, in this paper we have shown that those linear matrix inequalities provide not only necessary but also sufficient conditions. This leads to a more comprehensive solution concept since it can be adopted in other sensor design settings in a straightforward way in order to obtain an equivalent tractable optimization problem, e.g., for the settings over imperfect measurements as we did here, infinite horizon as shown in \cite{ref:Sayin19d} (based on the results of this paper), and several others.

\section{Proof of Lemma \ref{lem:linear}}\label{app:linear}

Based on \eqref{eq:optU} and \eqref{eq:etaStar}, we obtain \eqref{eq:short} through some algebra as detailed below. We focus on $i)$ part of the cost induced by the controller of the system, e.g., \eqref{eq:first}, and $ii)$ part of the cost induced by the attacker, e.g., \eqref{eq:second}, separately. 

{\em Part-$i)$} For notational brevity, let us first introduce the following matrices
\begin{align*}
&\Phi^{\omega} := \begin{bmatrix} 
I & K_{\kappa}^{\omega}B & K_{\kappa}^{\omega} AB & \cdots & K_{\kappa}^{\omega}A^{\kappa-2}B \\
& I & K_{\kappa-1}^{\omega}B & \cdots & K_{\kappa-1}^{\omega}A^{\kappa-3}B\\ 
& & I & \cdots & K_{\kappa -2}^{\omega}A^{\kappa-4}B\\
& & & \ddots & \vdots\\
& & & & I
\end{bmatrix}\\
&K^{\omega} := \begin{bmatrix} K_{\kappa}^{\omega} & & \\ & \ddots & \\ & & K_1^{\omega}\end{bmatrix},\;\Delta^{\omega} := \begin{bmatrix} \Delta_{\kappa}^{\omega} & & \\ & \ddots & \\ & & \Delta_1^{\omega}\end{bmatrix}.
\end{align*}
Then, the right hand side of \eqref{eq:control-free} can be written in a compact form as
\begin{equation*}
\|\Phi^{\omega}\ru + K^{\omega}\rx^o\|_{\Delta^{\omega}}^2 + \delta^{\omega}_0
\end{equation*}
in terms of the augmented vectors $\ru = \begin{bmatrix} \ru_{\kappa}' & \cdots & \ru_1'\end{bmatrix}'$ and $\rx^o = \begin{bmatrix} (\rx_{\kappa}^o)' & \cdots & (\rx_1^o)'\end{bmatrix}'$. Correspondingly, for type-$\omega$ \player{C}, we obtain
\begin{equation}\label{eq:uopt}
\ru^{\omega} = -(\Phi^{\omega})^{-1}K^{\omega}\rhx^o,
\end{equation}
where $\rhx^o := \begin{bmatrix} \E\{\rx_{\kappa}^o | \rs_{1:\kappa}\}' & \cdots & \E\{\rx_1^o|\rs_1\}' \end{bmatrix}'$ is the augmented vector of posteriors. Related to \eqref{eq:first}, this yields
\begin{align*}
\sum_{k=1}^{\kappa} \E\|\rx_{k+1}^{\omega_o}\|_{Q^{\omega_o}}^2 + \E\|\ru_k^{\omega_o}\|_{R^{\omega_o}}^2 &= \E\|K^{\omega_o}(\rx^o-\rhx^o)\|_{\Delta^{\omega_o}}^2 + \delta_o^{\omega_o}\nn\\
&= \trace\{HV^{\omega_o}\} + \xi^{\omega_o},
\end{align*}
where we define $\Sigma^o := \E\{\rx^o(\rx^o)'\}$,
\begin{align}
&V^{\omega_o} := -(K^{\omega_o})'\Delta^{\omega_o}K^{\omega_o},\label{eq:Xi1}\\
&\xi^{\omega_o} := \delta_o^{\omega_o} - \trace\{\Sigma^o \Xi^{\omega_o}\},\label{eq:xi1}
\end{align}
and $H:=\E\{\rhx^o (\rhx^o)'\}$, which follows since $\E\{\rx^o(\rhx^o)'\} = \E\{\rhx^o(\rhx^o)'\}$ due to the law of iterated expectations.

{\em Part-$ii)$} The state $\rx_k$ can be written in terms of control-free state $\rx_k^o$ and control inputs $\ru_{1:k}$ as follows:
\begin{equation}\label{eq:xo}
\rx_k = A\rx_{k-1}^o + \rw_{k-1} + \sum_{i=0}^{k-2} A^{i}B \ru_{k-i-1}.
\end{equation}
Let us define
\begin{equation*}
Z := \begin{bmatrix} B & AB & \cdots & A^{\kappa-1}B \\ & B & \cdots & A^{\kappa-2}B \\ & & \ddots & \vdots \\ & & & B\end{bmatrix}.
\end{equation*}
Then \eqref{eq:uopt} and \eqref{eq:xo} yield that for $\omega\in\Omega$, we have 
\begin{align}
\sum_{k=1}^{\kappa}\E\|\rx_{k+1}^{\omega}\|_{Q^{\omega_o}}^2 &= \E\|\bar{A}\rx^o + \rw - T^{\omega}\rhx^o\|_{\bar{Q}^{\omega_o}}^2\\
&= \trace\{HV^{\omega}\} + \xi^{\omega},\label{eq:w}
\end{align}
where $\rw := \begin{bmatrix} \rw_{\kappa}' & \cdots & \rw_1' \end{bmatrix}'$, $\bar{A}:=I_{\kappa}\otimes A$, $T^{\omega} := Z(\Phi^{\omega})^{-1}K^{\omega}$, $\bar{Q}^{\omega_o} = I_{\kappa} \otimes Q^{\omega_o}$, and for all $\omega\in\Omega$ we define
\begin{align}
&V^{\omega} := (T^{\omega})'\bar{Q}^{\omega_o}T^{\omega} -(T^{\omega})'\bar{Q}^{\omega_o}\bar{A} - \bar{A}'\bar{Q}^{\omega_o}T^{\omega},\label{eq:Xi2}\\
&\xi^{\omega} := \trace\{\Sigma^o \bar{A}' \bar{Q}^{\omega_o}\bar{A}\} + \kappa \trace\{\Sigma_w Q^{\omega_o}\}.\label{eq:xi2}
\end{align}
Note that \eqref{eq:w} follows since $\{\rw_k\}$ is a white noise and $T^{\omega}$ turns out to be an upper triangular (block) matrix.

Combining Parts $i)$ and $ii)$ together, we obtain that \player{S} faces the following problem: 
\begin{equation*}
\sum_{\omega\in\bOmega} p_{\omega}(\trace\{HV^{\omega}\} + \xi^{\omega}),
\end{equation*}
where $V^{\omega}$ and $\xi^{\omega}$ for $\omega\in\bOmega$ are as described in \eqref{eq:Xi1}, \eqref{eq:xi1}, \eqref{eq:Xi2}, and \eqref{eq:xi2}. Based on the definition of $H$, it can be shown that 
\begin{equation*}
H = \begin{bmatrix} H_{\kappa} & AH_{\kappa-1} & \cdots & A^{\kappa-1}H_1 \\ H_{\kappa-1}A' & H_{\kappa-1} & &A^{\kappa-2}H_1 \\ 
\vdots & & \ddots & \vdots \\ H_1(A^{\kappa-1})' & H_1 (A^{\kappa-2})' & \cdots & H_1 \end{bmatrix}.
\end{equation*}
Therefore, the corresponding $\Xi_k^{\omega}\in\universe{S}^m$ in \eqref{eq:short} is defined by
\begin{equation}\label{eq:V}
\Xi_k^{\omega} := V_{k,k}^{\omega} + \sum_{l = k+1}^{\kappa} V_{k,l}^{\omega} A^{l-k} + (A^{l-k})' V_{l,k}^{\omega},
\end{equation}
where $V_{k,l}^{\omega}\in\R^{m\times m}$ is an $m\times m$ block of $V^{\omega}$, with indexing from the right-bottom to the left-top. 

\section{Proof of Proposition \ref{prop:suff}}\label{app:suff}

The necessity condition has been shown in \cite[Lemma 3]{ref:Sayin17b}. 

In order to show the sufficiency of the condition, suppose that a collection of positive semi-definite matrices $S_{1:\kappa}$ satisfying \eqref{eq:constraint} is given. Then $S_1\in\universe{S}^m$ satisfies
\begin{equation}\label{eq:subConst1}
\Sigma_1^o \succeq S_1 \succeq O.
\end{equation} 
Note that $\Sigma_1^o\succeq O$ can be singular. Therefore let $\Sigma_1^o = \bU_1 \begin{bmatrix} \bLambda_{1} & O \\ O & O\end{bmatrix}\bU_1'$ be the eigen-decomposition such that $\bLambda_{1} \in\universe{S}_{++}^{t_1}$ and $t_1:=\rank\{\Sigma_1^o\}$. When we multiply the terms in \eqref{eq:subConst1} from right by the unitary matrix $\bU_1$ and from left by the transpose of the unitary matrix, i.e., $\bU_1'$, we obtain
\begin{equation*}
\begin{bmatrix} \bLambda_{1} & O \\ O & O\end{bmatrix} \succeq \bU_1' S_1 \bU_1 \succeq O,
\end{equation*}
which implies that
\begin{equation}
\begin{bmatrix} \bLambda_{1} & O \\ O & O\end{bmatrix} - \begin{bmatrix} M_{1,1} & M_{1,2} \\ M_{2,1} & M_{2,2} \end{bmatrix} \succeq O,\label{eq:this}
\end{equation}
where we let $\bU_1' S_1 \bU_1 = \begin{bmatrix} M_{1,1} & M_{1,2} \\ M_{2,1} & M_{2,2} \end{bmatrix}$ be the corresponding partitioning, e.g., $M_{1,1} \in\universe{S}^{t_1}$. 

Since $\bU_1' S_1 \bU_1 \succeq O$, we have $M_{2,2}\succeq O$  \cite[Observation 7.1.2]{ref:Horn85}. However, the bottom-right block of the positive semi-definite matrix (the whole term) on the left-hand-side of the inequality \eqref{eq:this}, i.e., $-M_{2,2}$, must also be a positive semi-definite matrix, which implies $O\succeq M_{2,2}$. Therefore we can conclude that $M_{2,2} = O$.

Next we invoke \cite[Lemma 3]{ref:Sayin18e} yielding $M_{1,2} = M_{2,1}' = O$. Therefore \eqref{eq:this} can be written as
\begin{equation}
\begin{bmatrix} \bLambda_{1} & O \\ O & O\end{bmatrix} - \begin{bmatrix} M_{1,1} & O \\ O & O \end{bmatrix} \succeq O.\label{eq:that}
\end{equation}
We define $T_1 := \bLambda_1^{-1/2}M_{1,1}\bLambda_1^{-1/2}$, let $T_1 = U_1\Lambda_1U_1'$ be its eigen-decomposition, and  let $\lambda_{1,1},\ldots,\lambda_{1,t_1}$ be its eigenvalues. Then \eqref{eq:that} yields that $I_{t_1}\succeq T_1 \succeq O_{t_1}$ and correspondingly $I_{t_1} \succeq \Lambda_1 \succeq O$, which implies that $\lambda_{1,i}\in[0,1]$ for all $i=1,\ldots,t_1$. 

Since $M_{1,1} = \bLambda_{1}^{1/2} T_1 \bLambda_{1}^{1/2}$, $M_{1,2}=M_{2,1}=O$, and $M_{2,2}=O$, $S_1$ can be written as
\begin{equation}\label{eq:ST}
S_1 = \bU_1 \begin{bmatrix} \bLambda_{1}^{1/2} T_1 \bLambda_{1}^{1/2} & O \\ O & O \end{bmatrix}\bU_1'.
\end{equation}
Since $\bLambda_1\in\universe{S}_{++}^{t_1}$ is not a singular matrix, \eqref{eq:ST} yields that there exists a bijective relation between $S_1\in\AS^m$ and $T_1\in\AS^{t_1}$, i.e., given $S_1$, we can compute $T_1$ and vice versa. Therefore, we can just focus on $T_1$ instead of $S_1$. To this end, consider a signaling strategy $\rs_1 = L_1'\rx_1 + \rn_1$, where $\rn_1\sim\N(0,\Theta_1)$. Then, the covariance matrix $H_1$ is given by
\begin{equation}\label{eq:HH}
H_1 = \Sigma_1^oL_1(L_1'\Sigma_1^oL_1 + \Theta_1)^{\dagger}L_1'\Sigma_1^o.
\end{equation}
Note that we can set $L_1\in\R^{m\times m}$ and $\Theta_1\in\universe{S}_+^m$ arbitrarily. Given the eigenvalues of $T_1$; it can be verified that if we set $L_1 = \bU_1 \begin{bmatrix} \bLambda^{-1/2}_1U_1\Lambda_1^o & O \\ O & O \end{bmatrix}$ and $\Theta_1 \succeq O$ such that
\begin{align*}
&\Lambda_1^o = \diag\{\lambda_{1,1}^o,\ldots,\lambda_{1,t_1}^o\},\\
&\Theta_1 = \diag\{\theta_{1,1}^2,\ldots,\theta_{1,t_1}^2,0,\ldots,0\}
\end{align*}
and the entries $\{\lambda_{1,i}^o,\theta_{1,i}^2\}$ satisfy
\begin{equation}\label{eq:ooo}
\frac{(\lambda_{1,i}^o)^2}{(\lambda_{1,i}^o)^2 + \theta_{1,i}^2} = \lambda_{1,i}, \mbox{ for } i=1,\ldots,t_1,
\end{equation}
then we obtain $H_1 = S_1$ exactly. Particularly, for such $L_{1,1}$ and $\Theta_1$, eigen-decomposition of $\Sigma_1^o$ yields that $L_1'\Sigma_1L_1$ can be written as
\begin{align*}
\begin{bmatrix} \Lambda_1^o \cancel{U_1'} \cancel{\bLambda_1^{-1/2}} & O \\ O & O \end{bmatrix}\cancel{\bU_1'}\cancel{\bU_1} \begin{bmatrix} \cancel{\bLambda_{1}} & O \\ O & O\end{bmatrix}&\cancel{\bU_1'}\cancel{\bU_1} \begin{bmatrix} \cancel{\bLambda^{-1/2}_1}\cancel{U_1}\Lambda_1^o & O \\ O & O \end{bmatrix}\nn\\
&=\begin{bmatrix}(\bLambda_1^o)^2 & O \\ O & O \end{bmatrix}
\end{align*}
since unitary matrices satisfy $\bU_1\bU_1' = \bU_1'\bU_1 = I$. On the other hand, $\Sigma_1^oL_1$ can be written as
\begin{equation*}
\bU_1\begin{bmatrix} \bLambda_{1} & O \\ O & O\end{bmatrix}\cancel{\bU_1'}\cancel{\bU_1} \begin{bmatrix} \bLambda^{-1/2}_1U_1\Lambda_1^o & O \\ O & O \end{bmatrix} = \bU_1\begin{bmatrix} \bLambda_1^{1/2}U_1\Lambda_1^{o} & O \\ O & O \end{bmatrix}.\nn
\end{equation*}
Therefore $\Sigma_1^oL_1(L_1'\Sigma_1^oL_1 + \Theta_1)^{\dagger}L_1'\Sigma_1^o$ can be written as
\begin{align}
\bU_1\begin{bmatrix} \bLambda_1^{1/2}U_1\Lambda_1^{o} & O \\ O & O \end{bmatrix}&\begin{bmatrix}(\bLambda_1^o)^2 + \Theta_1 & O \\ O & O \end{bmatrix}^{\dagger} \begin{bmatrix} \Lambda_1^{o}U_1'\Lambda_1^{1/2} & O \\ O & O \end{bmatrix}\bU_1'\nn\\
&=\bU_1\begin{bmatrix} \bLambda_1^{1/2}U_1\Lambda_1  U_1'\Lambda_1^{1/2} & O \\ O & O \end{bmatrix}\bU_1',\label{eq:sss}
\end{align}
which follows from \eqref{eq:ooo}. Recall that $T_1 = U_1\Lambda_1U_1'$. Therefore \eqref{eq:sss} is equivalent to \eqref{eq:ST}, which verifies the claim. Note also that there always exist $\{\lambda_{1,i}^o,\theta_{1,i}^2\}$ satisfying \eqref{eq:ooo} since $\lambda_{1,i}\in[0,1]$ for all $i=1,\ldots,t_1$.

We have shown that given $S_1\in\universe{S}^m$ satisfying \eqref{eq:subConst1}, we can select a signaling strategy such that $H_1 = S_1$ exactly. Next, by following similar lines, we compute the associated signaling strategies for $k>1$ under the assumption that we have obtained them up to $k-1$. 

Suppose that $H_j = S_j$ for $j<k$. Then, $S_k\in\AS^m$ satisfies
\begin{equation*}
\Sigma^o_k \succeq S_k \succeq AS_{k-1}A',
\end{equation*}
which is equivalent to
\begin{equation}\label{eq:asd}
\Sigma^o_k - AH_{k-1}A' \succeq S_k - AH_{k-1}A' \succeq O.
\end{equation}
Correspondingly, $\Sigma_k^o - AH_{k-1}A'\succeq O$ can be singular. Let $\Sigma_k^o - AH_{k-1}A' = \bU_k \begin{bmatrix} \bLambda_k & O \\ O & O\end{bmatrix}\bU_k'$ be the eigen-decomposition such that $\bLambda_k \in\universe{S}_{++}^{t_k}$ and $t_k := \rank\{\Sigma_k^o - AH_{k-1}A'\}$. When we multiply the terms in \eqref{eq:asd} from right by the unitary matrix $\bU_k$ and from left by the transpose of the unitary matrix, i.e., $\bU_k'$, we obtain
\begin{equation}\label{eq:asd3}
\begin{bmatrix} \bLambda_k & O \\ O & O \end{bmatrix} \succeq \bU_k'(S_k-AH_{k-1}A')\bU_k \succeq O.
\end{equation}
Following the same reasons for the case $k=1$, \cite[Lemma 3]{ref:Sayin18e} yields that there exists a symmetric matrix $T_k\in\AS^{t_k}$ such that
\begin{equation}\label{eq:asd2}
S_k = AH_{k-1}A' + \bU_k \begin{bmatrix} \bLambda_k^{1/2} T_k \bLambda_k^{1/2} & O \\ O & O \end{bmatrix}\bU_k'
\end{equation}
and there exists a bijective relation between $T_k$ and $S_k-AH_{k-1}A'$. Similarly, \eqref{eq:asd3} and \eqref{eq:asd2} yield that  
$I_{t_k}\succeq T_k \succeq O_{t_k}$, which implies that $T_k\in\AS^{t_k}$ has eigenvalues in the closed interval $[0,1]$. Let $T_k = U_k\Lambda_kU_k'$ be the eigen decomposition and $\lambda_{k,1},\ldots,\lambda_{k,t_k}\in[0,1]$ be the associated eigenvalues.

Furthermore, consider a signaling strategy $\rs_k = L_k'\rx_k + \rn_k$, where $\rn_k\sim\N(0,\Theta_k)$. Then, the covariance matrix $H_k$ is given by
\begin{align*}
H_k = A&H_{k-1}A' + (\Sigma_k^o - AH_{k-1}A')L_k\\
\times&(L_k'(\Sigma_k^o - AH_{k-1}A')L_k + \Theta_k)^{\dagger}L_k'(\Sigma_k^o - AH_{k-1}A'),
\end{align*}
which follows since
\begin{align*}
\cov\{\E\{\rx_k^o|\rs_{1:k}\}\} = \;&\cov\{\E\{\rx_k^o|\rs_{1:k-1}\}\} \nn\\
&+ \cov\{\E\{\rx_k^o | \rs_k - \E\{\rs_k|\rs_{1:k-1}\}\}\},
\end{align*}
due to the independence of the jointly Gaussian $\rs_{1:k-1}$ and $\rs_k-\E\{\rs_k|\rs_{1:k-1}\}$. We can again set $L_k\in\R^{m\times m}$ and $\Theta_k\in\universe{S}_+^m$ arbitrarily. Given the eigenvalues of $T_k$, it can be verified that if we set $L_k =  \bU_k\begin{bmatrix} \bLambda_k^{-1/2}U_k\Lambda_k^o & O \\ O & O \end{bmatrix}$ and $\Theta_k\succeq O$ such that
\begin{align*}
&\Lambda_k^o = \diag\{\lambda_{k,1}^o,\ldots,\lambda_{k,t_k}^o\},\\
&\Theta_k = \diag\{\theta_{k,1}^2,\ldots,\theta_{k,t_k}^2,0,\ldots,0\}
\end{align*}
and the entries $\{\lambda_{k,i}^o,\theta_{k,i}^2\}$ satisfy
\begin{equation*}
\frac{(\lambda_{k,i}^o)^2}{(\lambda_{k,i}^o)^2 + \theta_{k,i}^2} = \lambda_{k,i}, \mbox{ for } i=1,\ldots,t_k,
\end{equation*}
then we would obtain $H_k = S_k$ exactly. Therefore, by induction, we conclude that for any $S_{1:\kappa}$ satisfying \eqref{eq:constraint}, there exists a certain signaling strategy $\eta\in\Eta$ such that $H_k = S_k$ for all $k=1,\ldots,\kappa$.
 
\section{Proof of Proposition \ref{prop:existence}}\label{app:existence}

Since the objective function in \eqref{eq:maxmin} is continuous in the optimization arguments, and the constraint sets are decoupled and compact, the extreme value theorem and maximum theorem (showing the continuity of parametric maximization under certain conditions \cite{ref:Ok07}) yields the existence of a solution to \eqref{eq:maxmin}. 

\section{Proof of Theorem \ref{theorem:compute}}

Based on the existence result in Proposition \ref{prop:existence}, suppose that $S^*$ solves \eqref{eq:maxmin} and $p^*$ is the maximizer of \eqref{eq:inner} for $S^*$. Since $p^*\in\Delta(\bOmega)$, there must be at least one type with positive weight. For example, suppose positive weight for the type $\omega\in\bOmega$, i.e., $p_{\omega}>0$. This implies that 
\begin{equation*}
\xi^{\omega} + \trace\{S^*\Xi^{\omega}\} \geq \xi^{\omega'} + \trace\{S^*\Xi^{\omega'}\}\;\forall \omega'\in\bOmega
\end{equation*}
since $\xi^{\omega} + \trace\{S^*\Xi^{\omega}\} = \max_{\omega'\in\bOmega} \left(\xi^{\omega'} + \trace\{S^*\Xi^{\omega'}\}\right)$ by \eqref{eq:postar}. Furthermore, this also implies that
\begin{equation*}
\xi^{\omega}+ \trace\{S^*\Xi^{\omega}\} = \sum_{\omega'\in\bOmega}p_{\omega'}^{*}\left(\xi^{\omega'} + \trace\{S^*\Xi^{\omega'}\}\right)
\end{equation*}
since $\xi^{\omega} + \trace\{S^*\Xi^{\omega}\} = \xi^{\omega'} + \trace\{S^*\Xi^{\omega'}\}$ if $p_{\omega'}^*>0$. These necessary conditions yield that \eqref{eq:maxmin} is equivalent to
\begin{align*}
\min_{S\in\Psi} &\; \left[\xi^{\omega} + \trace\{S\Xi^{\omega}\}\right]\nn\\
\mathrm{s.t. }&\; \trace\{S(\Xi^{\omega}-\Xi^{\omega'})\} \geq \xi^{\omega'} - \xi^{\omega} \;\forall \omega'\in\bOmega 
\end{align*}
which is an SDP isolated from the distribution over the extended type set. Therefore, by searching over the extended type set $\bOmega$, we can compute the minimum of \eqref{eq:maxmin}, which is the minimum over $\bOmega$. Once the minimum value is computed, $S^*$ can be computed according to the corresponding type, i.e., \eqref{eq:Sstarj}.

\section{Proof of Lemma \ref{lem:measurements}}\label{app:measurements}

For a signaling strategy as described in \eqref{eq:teta}, we would have
\begin{align*}
\E\{&\rx_k^o|\eta_1(\ry_1),\ldots,\eta_k(\ry_{1:k})\} \nn\\
&= \E\{\rx_k^o| \E\{\rx_1^o|\teta_1(\ry_1)\},\ldots,\E\{\rx_k^o|\teta_1(\ry_1),\ldots,\teta_k(\ry_{1:k})\}\}\nn\\
&=\E\{\rx_k^o|\teta_1(\ry_1),\ldots,\teta_k(\ry_{1:k})\},
\end{align*}
where the last line follows since $\E\{\rx_l^o|\teta_1(\ry_1),\ldots,\teta_l(\ry_{1:l})\}$, for $l\leq k$, is just a measurable function of $\{\teta_1(\ry_1),\ldots,\teta_l(\ry_{1:l})\}\}$ while $\E\{\rx_k^o|\teta_1(\ry_1),\ldots,\teta_k(\ry_{1:k})\}$ is already conditioned on $\{\teta_1(\ry_1),\ldots,\teta_k(\ry_{1:k})\}$. This yields that they both would lead to the same posterior. Recall that the best reaction of \player{C} is linear function of $\E\{\rx_{k}^o|\rs_{1:k}\}$ as described in \eqref{eq:optU}. Therefore they both would lead to the same control inputs almost surely and therefore the same cost for \player{S}.

\section{Closed-Form Expressions for Auxiliary Parameters Under Imperfect Measurements}\label{app:auxiliary}

Note that $\Sigma_k^y\in\AS^{nk}$, $A_k \in \R^{nk\times n(k-1)}$, and $D_k\in\R^{m\times nk}$ can be written as 
\begin{align*}
&\Sigma_k^y = \begin{bmatrix} O & I_k \otimes C\end{bmatrix}\Sigma^o \begin{bmatrix} O \\ I_k \otimes C' \end{bmatrix} + I_{k} \otimes \Sigma_v,\\
&A_k = \begin{bmatrix} \begin{bmatrix} O_{n\times (\kappa-k)n} & C & O_{n\times(k-1)n} \end{bmatrix}  \Sigma^o \begin{bmatrix} O \\ I_{k-1}\otimes C'\end{bmatrix} (\Sigma_{k-1}^y)^{\dagger} \\ I_{(k-1)n} \end{bmatrix},\\
&D_k = \begin{bmatrix} O_{m\times (\kappa-k)m} & I_m & O_{m\times(k-1)m} \end{bmatrix}  \Sigma^o \begin{bmatrix} O \\ I_k\otimes C'\end{bmatrix} (\Sigma_k^y)^{\dagger}.
\end{align*}

\bibliographystyle{IEEEtran}
\bibliography{ref}

\begin{IEEEbiographynophoto}{Muhammed O. Sayin} received the B.S. and M.S. degrees in electrical and electronics engineering from Bilkent University, Ankara, Turkey, in 2013 and 2015, respectively. He received the Ph.D. degree in electrical and computer engineering from the University of Illinois at Urbana-Champaign (UIUC) in 2019. He is currently a Postdoctoral Associate at Laboratory for Information and Decision Systems (LIDS) at MIT. His current research interests include dynamic games and decision theory, information design problems, and multi-agent systems.
\end{IEEEbiographynophoto}

\begin{IEEEbiographynophoto}{Tamer Ba\c{s}ar} (S'71-M'73-SM'79-F'83-LF'13)
is with the University of Illinois at Urbana-Champaign, where he holds the academic positions of  Swanlund Endowed Chair; Center for Advanced Study Professor of  Electrical and Computer Engineering; Research Professor at the Coordinated Science Laboratory; and Research Professor  at the Information Trust Institute. He is also the Director of the Center for Advanced Study.

He received B.S.E.E. from Robert College, Istanbul,
and M.S., M.Phil, and Ph.D. from Yale University. He is a member of the US National Academy
of Engineering,  member of the  European Academy of Sciences, and Fellow of IEEE, IFAC (International Federation of Automatic Control) and SIAM (Society for Industrial and Applied Mathematics), and has served as president of IEEE CSS (Control Systems  Society), ISDG (International Society of Dynamic Games), and AACC (American Automatic Control Council). He has received several awards and recognitions over the years, including the highest awards of IEEE CSS, IFAC, AACC, and ISDG, the IEEE Control Systems Award, and a number of international honorary doctorates and professorships. He has over 900 publications in systems, control, communications, and dynamic games, including books on non-cooperative dynamic game theory, robust control, network security, wireless and communication networks, and stochastic networked control. He was the Editor-in-Chief of Automatica between 2004 and 2014, and is currently  editor of several book series. His current research interests include stochastic teams, games, and networks; distributed algorithms; security; and cyber-physical systems.
\end{IEEEbiographynophoto}
\vfill

\end{document}